\documentclass[journal,comsoc]{IEEEtran}
\usepackage{algorithmic}
\usepackage{algorithm}
\usepackage{amsmath}
\usepackage{ifpdf}
\usepackage{graphicx}
\usepackage{graphics}
\usepackage{subcaption}
\usepackage{balance}
\usepackage{flushend}
\usepackage{cite}
\usepackage{cases}

\begin{document}
	
\title{Towards Skewness and Balancing of RPL Trees for the Internet of Things}

\author{Duc-Lam Nguyen,~\IEEEmembership{Student Member,~IEEE,}
	and Chong-Kwon Kim,~\IEEEmembership{Senior Member,~IEEE.}% <-this % stops a space

	\thanks{This work was supported by the National Research Foundation of Korea (NRF) Grant funded by the Korean Government (MSIP) (No.2016R1A5A1012966, and No. NRF-2015R1A2A1A01007400) and by Institute for Information \& communications Technology Promotion(IITP) grant funded by the Korea government(MSIT) (No.2015-0-00557, Resilient/Fault-Tolerant Autonomic Networking Based on Physicality, Relationship and Service Semantic of IoT Devices). Also, this work was partly supported by the Institute for Industrial Systems Innovation of Seoul National University \textit{(Corresponding author: Duc-Lam Nguyen, Chong-Kwon Kim)}.}%}
	
	\thanks{Duc-Lam Nguyen is with Social Computer \& Networks Laboratory, Department of Computer Science and Engineering, College of Engineering, Seoul National University, 1 Gwanak-ro, Gwanak-gu, Seoul 08826, Republic of Korea. Email: lam@popeye.snu.ac.kr.}% <-this % stops a space
	
	\thanks{Chong-Kwon Kim is with the Department of Computer Science and Engineering, College of Engineering, Seoul National University, 1 Gwanak-ro, Gwanak-gu, Seoul 08826, Republic of Korea. Email: ckim@popeye.snu.ac.kr.}% <-this % stops a space
%	\thanks{Email: lam@popeye.snu.ac.kr, ckim@snu.ac.kr.}	
}
%\markboth{IEEE Transactions on Green Communications and Networking,~Vol.~XX, No.~XX, XXX~2019}%
{}
\maketitle

\begin{abstract} 
In many application areas such as large-scale disaster detection, IoT networks connote the characteristics of LLN (Low power and Lossy Network). With few exceptions, prior work on RPL(Routing Protocol for LLN), a standard routing protocol standardized in the IETF, has focused on the evaluation of various aspects of routing performances and control overheads. In this paper, we address the problem of DODAG (Destination Oriented Directed Acyclic Graph) created by direct application of RPL. We first evaluate the skewness of DODAG both via numerical simulations and via actual large-scale testbed. RPL secures its flexibility and wide applicability by allowing the adoption of implementer-specific rank definitions and parent selection criteria. In addition to the metrics used in ContikiRPL and TinyRPL, the two most widely used open source implementations, we evaluated the skewness of RPL trees generated by applying various routing metrics. Performance analysis results show that RPL trees suffer from severe skewness regardless of routing metrics in both randomly generated networks and in real-world networks. We propose a novel routing protocol that may improve the balance of RPL trees. Rigorous performance analysis based on computer simulations shows that our algorithm improves the tree balance significantly. 
\end{abstract}


\section{Introduction}
\subsection{Overview}
With great progress and development made in information and communication technology, Internet of Things (IoT) and Machine-to-Machine (M2M)\cite{iot} have merged to provide ubiquitous communication of smart embedded devices, so that retrieving real-time information can become possible\cite{in40}\cite{gungor}. Due to the great potential brought by M2M and IoT communication, they are being considered as the evolutionary change in the field of wireless communications. A potential large number of nodes is able to establish low-power short-range wireless links, thus forming a capillary network infrastructure that can be connected to the global Internet\cite{50b}. A new class of multi-hop wireless sensor network has emerged that is generally characterized by a resource constrained failure-prone architecture and subsequently has given rise to new challenges to provide robustness and resilience \cite{ieee802154,wirelesshart}. These types of WSN are used in natural disaster monitoring, surveillance and industrial management where a certain reliability should be guaranteed while providing robustness in the presence of harsh surroundings\cite{ieee802154,wirelesshart,rpl,culler, icoin, anci,ko,lam}. The analysis of the different application scenarios has demonstrated that the routing protocol for LLNs should be able to cope with resource-constraint, quality of service and scalability issues. Several routing protocols have been introduced to figure out these issues such as AODV\cite{aodv}, Collection Tree Protocol\cite{ctp}, and LOAD\cite{load}.%

In order to achieve reliable and energy efficient data collection, the Internet Engineering Task Force (IETF) has proposed RPL \cite{rpl} as an IoT routing standard for IPv6 Low-Power and Lossy Networks (LLNs). RPL is an oriented distance vector routing protocol that allows users to establish logical routing topology known as a Destination-Oriented Directed Acyclic Graph (DODAG) structure, meaning that each node may have one or more than one parent towards the sink. RPL is designed to meet the different requirements of 6LoWPANs, it guarantees a fast network establishment which allows the efficient monitoring of critical applications. RPL is one of the most promising routing solution for a wide range of network types as well as industrial applications such as Smart Gird\cite{smartgrid}, Building Automation\cite{bldauto}, Home Automation\cite{Home}, and Advanced Metering Infrastructure (AMI)\cite{ami}.%

\subsection{Motivation}

Recently, RPL provisions several robust features such as self-healing, loop-free network, and exiguous delay. However, the load balancing has been considered as a weakness in the RPL standard. The routing protocol for LLNs should be lightweight, specifically in LLNs in which nodes are equipped with highly resource-constraints and featured short range communication abilities. Thus, high protocol overhead associated with path maintenance and discovery might drain resources quickly and interfere with data transmission. 

On the one hand, depending on the specific requirements, different routing metrics and constraints\cite{metrics} can be adopted such as hop-count\cite{of0}, latency, energy consumption or expected transmission count (ETX)\cite{etx}. Routing path construction relying solely on a single pairwise transmission quality metric may not be able to capture the real communication scenarios. The sizes of the networks necessitate the need to communicate over multiple hops requiring higher layer protocol support. Reliable and efficient of communications in large LLNs has yet to be sufficiently addressed\cite{50b}. Potential future applications will inevitably require the need to communicate beyond the range of sinks and require larger networks than that are supported recently. The fact that large-scale LLNs are not common is likely due to lack of support from current protocols and approaches, so motivating our research.   
 
On the other hand, LLNs are resource-constraint networks, it is a requirement of RPL to be energy efficiency. So that, RPL needs to balance not only the traffic load but also the number of connections of each node to provide fair energy consumption among nodes. RPL is designed for LLNs and performs routing in a distributed way, however, the load balancing feature is missing in RPL. Without load balancing the data traffic and the distribution of wireless sensor nodes in LLNs may result in significant unbalance for those nodes that have more neighbor nodes than others. As mentioned above, in large-scale networks, the nodes close to the gateway often handle heavy traffic load even others generate lightly traffic load. Thus, this results in gaps and holes in the whole network and causes the disconnected of the network connection. It leads to RPL needs to address the load imbalance problem. 

\subsection{Key Idea}

The key idea of this study is that we investigate the topology construction of RPL not only using casual metrics as standard RPL but also exploiting the skewness and balancing to apply the combination of metrics. We achieved the balance and the stability by taking into account the size of DODAG subtrees for selecting each parent candidate in the parent selection procedure. We defined a new specific metric representing for the influence of parent candidates to new joining nodes for routing procedure. In detail, a node willing to join DODAG should consider both the link quality with parent candidates and the influence of parent candidates to joining node, so the stability and balancing of routing path are guaranteed and reliable. The detail is described in Section III.  

\begin{figure}[t!]
	\centering
	\includegraphics[width=0.9\linewidth]{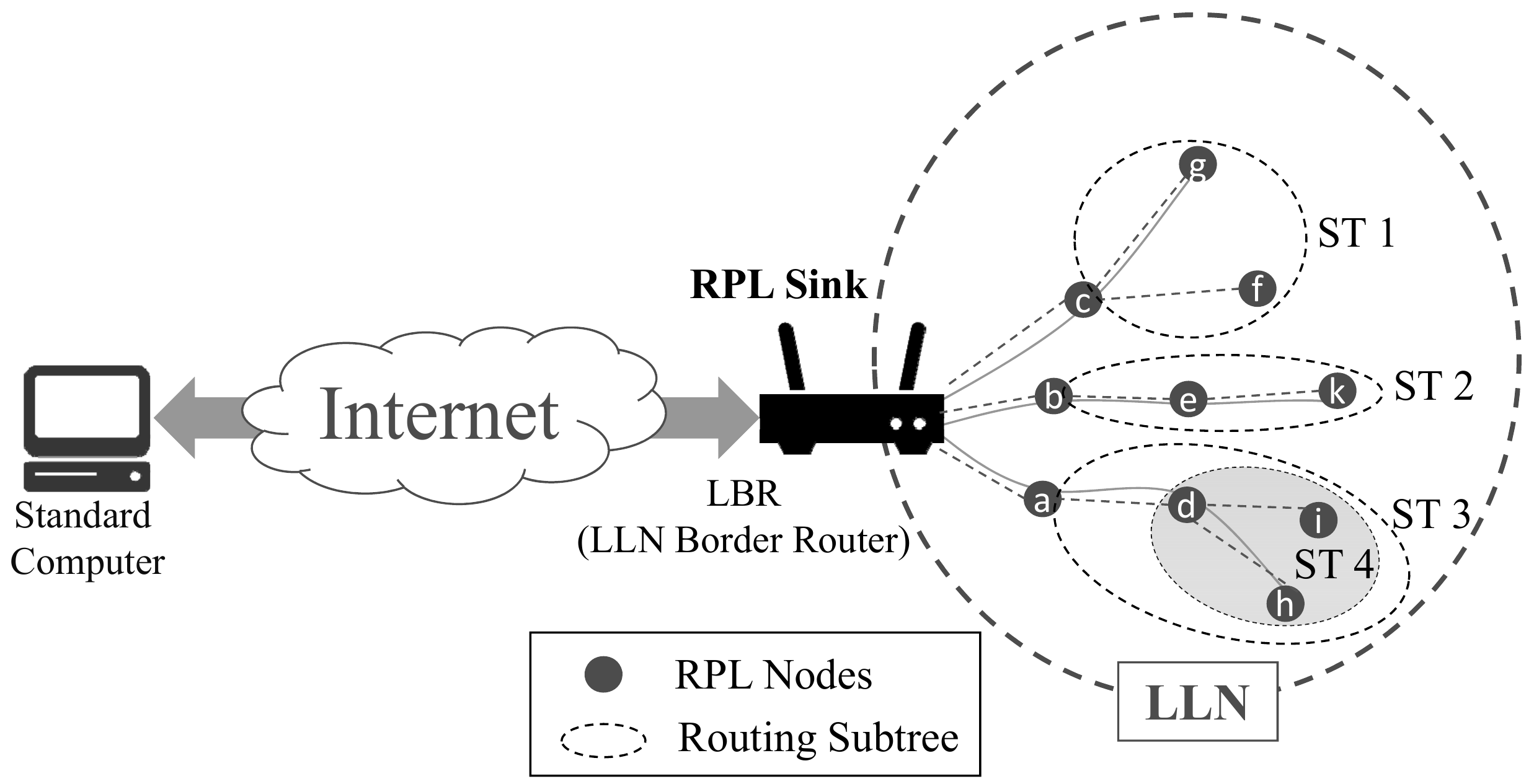}
	\caption[]{\textbf{An example of IoT multi-hop LLN}. The LLN is connected to Wide Area Network (WAN) which might be public global Internet via LBR (LLN Border Router). In this example, the LLN includes one sink (DODAG root), 10 source nodes, and several subtrees namely ST1 ($ST_{(c)}$), ST2 $ST_{(b)}$, ST3 ($ST_{(a)}$), ST4 ($ST_{(d)}$) and so on.}
	\label{fig:example}
\end{figure}

\subsection{Contributions}
With the aforementioned motivations and ideas, we propose SB-RPL, standing for \underline{\textbf{S}}kewness and \underline{\textbf{B}}alancing of \underline{\textbf{R}}PL Trees for IoT networks, a new extension of RPL that provides enhanced support for large-scale network and incorporates the load balance mechanism into RPL. SB-RPL is able to effectively increase the end-to-end reliability as well as the network balance.

We implemented SB-RPL in ContikiOS \cite{contiki} and conducted extensive numerical simulations using Contiki Cooja simulator and experiments using actual large-scale testbed FIT-IoT-Lab \cite{iotlab} with 100 nodes Arm-M3-Cortex\cite{arm}. In total, our evaluation based on around hundreds of individual simulations and experiments, the duration is from one to two hour per experiment\footnote{In terms of simulation, with the large-scale networks, one hour in simulator can equivalent to several hours in real-life.}. Our evaluation shows that SB-RPL improves not only skewness and balancing of RPL trees but also reliability and end-to-end delay significantly in comparison with existing RPL studies in both practical experiments and simulations.

The main contributions of this paper can be summarized as follows: 
\begin{enumerate}
	\item We proposed SB-RPL, the first work that investigates skewness and balancing and evaluates the performance in both Cooja simulation environment and practical large-scale FIT-IoT-Lab platform of Lille, France. SB-RPL exploits the combination of multiple metrics and skewness for routing efficiency in RPL DODAG (Section III.D).
	\item SB-RPL uses extended control message structures based on the standard structure defined in the specification of RPL. This makes sure that our proposed scheme SB-RPL is interoperable with standard RPL, thus LLN devices using standard RPL or SB-RPL can operate together seamlessly in a hybrid environment (Section III.C). 
	\item Our proposed scheme SB-RPL not only improves the skewness and balancing among subtrees in a DODAG but also supports adaptivity and mobility of the network without requiring specific statical assumptions on the Objective Functions. This factor is convenient for implementation in the actual environment because Objective Functions in IoT applications can be widely dissimilar. On the other hand, there is no any constraint on the designs of Objective Functions in the specification of RPL, it keeps opening for new researches (Section IV).     
	\item We implemented SB-RPL in ContikiOS which is an open source operating system for IoT and LLNs. Via extensive computer simulations using Contiki's network simulator and real-world experiments on the FIT IoT-LAB testbed, we proved that our proposed scheme significantly outperforms the existing methods in terms of reliability, adaptability to network balance of LLNs under various scenarios(Section IV).
\end{enumerate}

\subsection{Paper Organization}
The remainder of our paper is structured as follows: Section II provides prior works in this area. Our proposed scheme named SB-RPL is described in Section III. In particular, it covers high level detail of RPL protocol and we describe our SB-RPL protocol more specific. Section IV presents results from the performance evaluation of SB-RPL and discusses issues that may have a significant impact on its behaviors. The detail of our evaluation method such as information of testbed, simulator, evaluation metrics are explained in detail. Finally, we conclude the paper in Section V.%

\section{Related Work}
In this section, we review earlier works on objective functions and load balancing problems in low-power wireless networks. 

Recently, RPL has received significant attention from academic and industrial communities. Although there is a lot of studies about new objective functions as well as performance evaluation of RPL\cite{wang, clausen,triphathi,gaddour,potsch,herberg,hui,khan}, still there is a lack of studies about load balancing in RPL over actual large-scale multi-hop LLNs. We review previous studies in two categories: (1) RPL Objective Functions, (2) Balance Routing protocols. 
%Finally, we conclude related work with pointing out the different of our proposed scheme with prior works. 

\subsubsection{RPL Objective Functions}
RPL standard does not force the use of any specific objective function or any specific metric keeping in open for research. There are various objective functions that are used in the RPL network, the two most important objective functions are the Minimum Rank with Hysteresis Objective Function (MRHOF) and Objective Function Zero (OF0). Several approaches were proposed in the literature attempting to develop objective functions for RPL \cite{of0},\cite{etx},\cite{todoli}. 
In order to handle congestion problems that occur in terms of heavy data transmission, the work in \cite{caof} introduced a congestion-aware objective function CA-OF which considered buffer occupancy as the routing factor. CA-OF showed an improvement of packet delivery ratio by avoiding congested nodes in routing paths in case of heavy data traffic. However, the routing stability is not considered in this research.

Iova et al. \cite{lova} proposed a new metric named Expected Lifetime and combined it in the calculation with  data traffic load and link reliability when estimating how long a node could stay before exhausting its own residual energy. The purpose of this method was to maximize the lifetime of most constrained nodes. However, the proposed method showed a high computational overhead that is unfeasible in LLNs environment. To address the limitations of network scalability, Songhua et al. \cite{songhua} proposed a Qos-aware fuzzy logic objective function. This objective function includes four metrics namely hop count, delay, ETX and battery level which can estimate the path quality using fuzzy logic techniques. Several studies introduced the different mechanism to optimize routing metrics efficiently and new objective functions for RPL to meet vary requirements in specific application environments \cite{qurpl},\cite{gonzizi}.%

\subsubsection{Balance Routing Protocols}
The RPL balancing problem has been investigated in several prior studies such as \cite{xliu,boubekeur,lodhi,nassiri}. 
In \cite{nassiri}, the authors proposed a new parent selection procedure. In which, generated parent set considers both the Received Signal Strength Indicator (RSSI) and residual energy, the proposed method selects probabilistically the parent node for every data transmission. The authors in \cite{qurpl} combined queue information with Objective Function Zero to enhance load-balancing of RPL routing under heavy traffic scenarios. Manually setting parameters, as suggested in QU-RPL, is challenging in dynamic and large-scale IoT networks and it is limited to OF0. To address the load imbalance of ORPL \cite{orpl}, Michel et al. \cite{michel} proposed ORPL-LB which achieved load balancing by using a sleep interval control mechanism and selective ACK transmission. Via various experiments, the authors proved that ORPL-LB has a better battery lifetime among nodes than standard RPL and ORPL. 

The authors \cite{xliu} proposed LB-RPL which improves load balancing of RPL by allowing a node to prioritize its parent candidates based on their queue utilization. The queue utilization information is collected from its neighbor nodes through DIO transmission. If congestion is detected, then the nodes delay the dissemination of routing information. M-RPL \cite{lodhi} detects traffic congestion problem by using RPL control messages and provides two preferred parent nodes for traffic distribution. The work in \cite{alabamo}, ALABAMO was proposed which supports MRHOF in load balancing capacity. With ALABAMO, RPL nodes consider parent selection process using both ETX and traffic load value. Through actual experiments, the authors demonstrated the improvement in load balancing of ALABAMO but they did not consider a duty cycling mechanism for evaluation and they assumed simply that fair relay burden can balance routing lifetime.%
\begin{table} [t!]
	\caption{Notations}
	\begin{tabular}{ p{1.5cm} p{6.5cm} }
		\hline
		\hline
		\textbf{Notations}  & \textbf{Descriptions}  \\ 
		\hline
		\hline
		$\mathcal{N}$ & Number of sensor nodes\\
		$\mathcal{L}$ & The set of all wireless links\\
		$\mathcal{S}$ & Sink of DODAG \\
		$\mathcal{R}_n(t)$ & Rank of node $n$ at timeslot $t$ \\		
		$c_{n,p}(t)$ & Logical Link-layer channel capacity between node $n$ and node $p$\\		
		$c^{max}$ & maximum channel capacity value\\
		$l_{n,p}(t)$ & Link characteristics between node $n$ and node $p$ \\
		$\mathcal{P}_{n,p}(t)$ & Node $p$ is preferred parent of node $n$ \\	
		%		$\mathcal{C}^max_n(t)$ & Maximum number of children node $n$ can maintain \\
		$\mathcal{RT}^S$ & Routing Subtree \\
		$N^S$ & Set of sensor nodes in a subtree \\
		$L^S$ & Set of direct links in a subtree \\ 
		$ST_{p}(t)$ & Subtree size of node $p$ at timeslot $t$\\
		$ST^{max}_{p}(t)$ & Maximum subtree size of node $p$ at timeslot $t$\\
		$ST^{min}_{p}(t)$ & Minimum subtree size of node $p$ at timeslot $t$\\
		$ST^{avr}_{p}(t)$ & Average subtree size of node $p$ at timeslot $t$\\		
		$NI_{n,p}(t)$ & Node Influence of potential parent $p$ to the new joining node $n$ at time $t$ \\
		$\mathcal{NB}_n(t)$ & Set of neighbors that node $n$ can communicate during timeslot $t$ \\
		
		$PRR_{n,p}(t)$ & Packet Reception Ratio between node $n$ and node $p$ \\
		
%		$\delta$ & Parent switch threshold \\
%	$C_n$ & Number of child nodes of node $n$ \\
		$\mathcal{M}1$ & Skewness metric 1 \\
		$\mathcal{M}2$ & Skewness metric 2 \\
		$\mathcal{M}3$ & Skewness metric 3 \\
		$\mathcal{M}4$ & Skewness metric 4 \\
		
		\hline		
	\end{tabular}
\end{table}
\section{SB-RPL Design}
In this section, we model RPL in detail and describe our proposed named SB-RPL, aiming to enhance the skewness and balancing of RPL DODAG as well as improve the performance of RPL in terms of reliability, end-to-end delay, and adaptivity to the dynamics of resource-constraint networks. %
 
\subsection{System Models}
The suggested SB-RPL approach is designed for LLN networks organized in a single DODAG. Thus, we consider a LLN as a set of multiple IoT devices. The LLN $\mathcal{G} = (\mathcal{N}, \mathcal{L})$ includes $\mathcal{N}$ standing for set of sensor nodes and $\mathcal{L}$ standing for set of direct links. The network operates in discrete time slots (e.g., seconds): t = \{0,1,2,3,...m\}. 

\subsubsection{Low-Power and Lossy Network (LLN)}
To model the unreliable and lossy wireless transmissions, we use packet reception ratio (PRR) over the wireless link from node $n$ to node $p$ as (n,p), $0 \leq PRR_{n,p}(t) \leq 1$. The PRR is as the probability of successfully transmitting a packet and then receiving an acknowledgment between node $n$ and node $p$ at timeslot $t$. For RPL, we can calculate $ETX$ (Expected Transmission Count) from PRR. $ETX$ is the measure for determining the total number of retransmissions required to successfully transmit data packet to next node with an acknowledge. Considering a given PRR value between node $n$ and node $p$ as $PRR_{n,p}(t)$, the corresponding ETX value $ETX_{n,p}(t)$ can be achieved as followed: 
\begin{eqnarray}
ETX_{n,p}(t) = \frac{1}{PRR_{n,p}(t)}
\end{eqnarray}
RPL smooths ETX using an exponential weighted moving average (EWMA) filter\cite{ewma} which is widely used method to update statistics such as average and standard deviation, making it robust to sudden changes in RPL DODAG. It updates ETX as:  
\begin{equation}
	ETX_{n,p}(\mathrm{new}) = \gamma ETX_{n,p}(\mathrm{current}) + (1-\gamma)ETX_{n,p}(\mathrm{packet})  
\end{equation} 
where $ETX_{n,p}(current)$ is the ETX metric that node $n$ currently has for its parent node $p$, and $ETX_{n,p}(new)$ is the ETX value obtained from the last single transmission from the child nodes. The default value of $\gamma$ is set to 0.1.

We define the logical link-layer channel capacity $c_{n,p}(t)$ of a wireless link from node $n$ to node $p$ at time slot $t$ as followed: 
\begin{equation}
c_{n,p}(t) = c^{\max}_{n,p}PRR_{n,p}(t) = \frac{c^{\max}_{n,p}}{ETX_{n,p}(t)}
\end{equation}
$c_{n,p}(t)$ presents the number of acknowledgment packets transmitted from node $n$ to node $p$ with timeslot $t$, $c^{\max}_{n,p} \geq c_{n,p}(t) \geq 0$ with $c^{max}(n,p)$ is the maximum value of $c_{n,p}(t), \forall t$. If capacity channel value $c_{n,p}(t)>0$, it means that node $n$ and node $p$ are in communication at time slot $t$; otherwise, they are not in communication at timeslot $t$. 

We denote that $\mathcal{NB}_{n}(t)$ is set of all neighbors that node $n$ can communicate during timeslot $t$, $\mathcal{NB}_{n}(t) \in \mathcal{N}$ : 
\begin{equation}
\mathcal{NB}_{n}(t) := \{p | c_{n,p}(t) > 0, c_{p,n}(t) > 0, p \in \mathcal{N} - \{n\}\}
\end{equation}

As the specification of original RPL, neighbor table have several main policies as followed: 

\textbf{NP1}: \textit{A node $n$ adds the neighbor $k$ if there is an indication that this is a better parent than the worst of the current parent.}

\textbf{NP2:} \textit{ When a node $n$ have empty neighbor table $\mathcal{NB}_{n}(t)$, it can always add new neighboring nodes to its neighbor table.}

\textbf{NP3:} \textit{Node $n$ will add node $k\in\mathcal{NB}_n(t)$ to its neighbor table if there is enough space for other children and send a DAO-ACK message. The nodes already in the table of node $n$ are not deleted except the lifetime timeout is expired.}

\textbf{NP4:} \textit{When node $n$ receives a DIS message of node $k$, if this DIS is a unicast transmission, node $n$ will add node $k$ to $\mathcal{NB}_n(t)$; otherwise, node $n$ ignores the DIS.}

The state of low-power and lossy network at a given timeslot $t>0$ can be presented as a directed and modeled as a time-varying weighted graph $\mathcal{G(N, L}, c(t))$ where $\mathcal{N}$ is the set of sensor nodes and $\mathcal{L}$ is all possible links for all nodes pairs in $\mathcal{N}$. 

Each RPL node n recognizes its neighbors by receiving DIO messages. Then, each node generates its own parent candidate set $P_n$ from its neighbor set $\mathcal{NB}_{n}(t)$ as followed:

\begin{equation}
\mathcal{P}_n(t) = \{p \in \mathcal{NB}_{n}(t)  | \mathcal{R}_n < \mathcal{R}_p, ETX_{n,p}(t) < \delta \}
\end{equation}%
where $\delta$ is a threshold to remove neighbors which are connected through unreliable links.

\subsubsection{RPL Objective Function}
RPL constructs a DODAG by using a specific Objective Function which defines a routing optimization objective that translates one or more metrics and constraints such as latency, minimizing energy consumption or ETX into a value called rank. RPL defines rank to indicate the routing distance from a node to sink, which is attached in DIO messages and used to parent selection procedure. In ContikiRPL,  MRHOF is used as default objective function where rank is computed based on ETX information.

The rank value of new device $n\in \mathcal{N}$  is determined by the following formula:
\begin{equation}
	\mathcal{R}_n(t) =\left\{\begin{IEEEeqnarraybox}[\relax][c]{l's}
	\min_{p\in \mathcal{NB}_n(t)}(l_{n,p}(t) + \mathcal{R}_p(t)) & $n \neq \mathcal{S} $  \\
	R_\mathcal{S} & $n=\mathcal{S}$
\end{IEEEeqnarraybox}\right.
\end{equation}
where $\mathcal{R}_n(t)$ is the rank of node $n$, $\mathcal{R}_p(t)$ is the rank of the potential parent DODAG node $p$ of the node $n$, and $l_{n,p(t)}$ is a function of the characteristics of node $p$ and of the link between node $n$ and node $p$ at timeslot $t$. $\mathcal{R}_\mathcal{S} \geq 0$  is rank of the sink, the smallest rank value in the DODAG architecture. Thus, when a new device $n$ willing to join DODAG, its rank is computed by searching a one-hop neighbor that gives the smallest sum of link characteristic $l_{n,p}(t)$ and neighbor rank $\mathcal{R}_p(t)$. In ContikiRPL, $l_{n,p}(t)$ represents for $ETX_{n,p}(t)$.

In the parent selection process, each node  selects its best parent $\mathcal{P'}_n$ from parent candidate set $\mathcal{P}_n$: 
\begin{equation}
P'_n(t) =  \arg \min_{p\in \mathcal{P}_n(t)} (\mathcal{R}_p(t))
\end{equation}

If the smallest path cost for paths through the candidate neighbors is smaller than the current path cost by less than a threshold, the node may continue to use the current preferred parent. This is considered as a hysteresis component of MRHOF objective function. Then a node may change its preferred parent if its information on parent candidates has been changed if:
\begin{equation}
	\mathcal{R}(P'_n) < \mathcal{R}(\mathcal{P}_n) + \sigma
\end{equation}
where $\sigma$ is a stability bound to mitigate unnecessarily and inefficiently the parent change, which is set to 96 by default. This $\sigma$ is the difference between ETX of the route through the preferred parent and the minimum-ETX route to trigger a new preferred parent procedure. Each RPL node selects a parent node which has a reliable link or minimum hop distance to the sink, regardless of traffic load and balancing for DODAG.

The RPL objective functions have several main properties as follows:
 
\textbf{P1}: \textit{If there is a change of $\mathcal{NB}_n(t)$ such as adding or removing entries or the entries are changed, then the node $n$ will eventually re-compute $\mathcal{R}_n(t)$ and re-select $P'_n(t)$.}

\textbf{P2}: \textit{When the node $n$ re-selects its preferred parent $P'_n(t)$ and $\mathcal{R}_n(t)$, a non-sink node adopts \textit{NULL} as $P'_n(t)$ if it also adopts infinite rank. There are the initial values of preferred parent and rank at the non-sink node when the node needs to (re)start.}

\textbf{P3}: \textit{Similarly, when the node $n$ re-selects its preferred parent $P'_n(t)$ and $\mathcal{R}_n(t)$, the sink adopt \textit{NULL} and \textit{MinHopRankIncrease}, respectively.}

\textbf{P4}: \textit{When node $n$ reselects $P'_n(t)$, the non-sink node will adopts \textit{NULL} and infinity, respectively. And if its $\mathcal{NB}_n(t)$ does not include any entry, for which $R_n(t)$ can be calculated by an objective function, and these following constraints should be satisfied: 
\begin{itemize}
	\item $R_{\mathcal{NB}_n(t)}$ $<$ infinity 
	\item $R_n(t)$ $\le$ $\mathcal{R}_{\min}(t)$ + \textit{MaxRankIncrease}
	\item  $R_n(t)$ $\ge$ $R_{\mathcal{NB}_n(t)}$ + \textit{MinHopRankIncrease}
	\item  Node $n$ and $\mathcal{NB}_n(t)$ are reachable. 
\end{itemize}
Otherwise, node $n$ will select a neighbor as a preferred parent.}
\textbf{P5}: \textit{The $R_n(t)$ and $\mathcal{P}_n(t)$ change only as a result of re-selection of the node's death and reset; otherwise they keep stably.}
\begin{figure}[t!]
	\centering
	\includegraphics[width=0.9\linewidth, height=0.18\textheight]{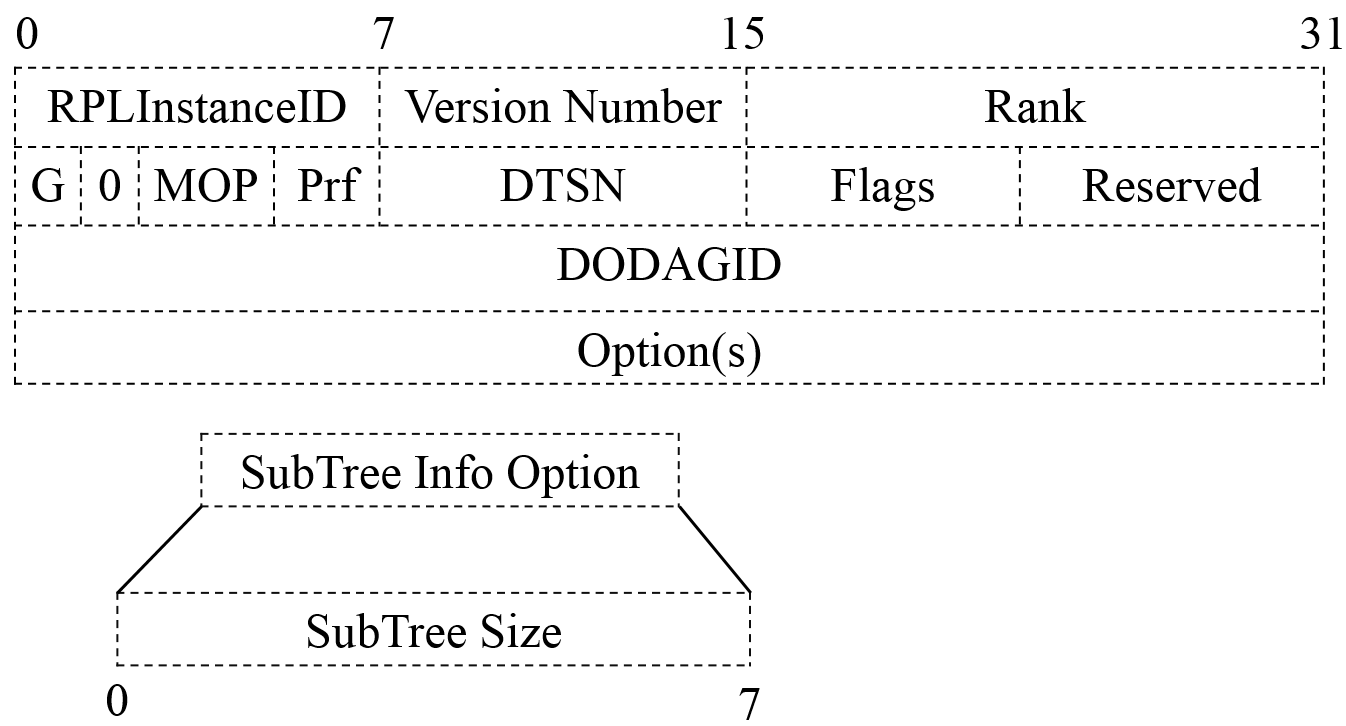}
	\caption{\textbf{RPL DIO control message  structure in SB-RPL}}
	\label{fig:diostructure}
\end{figure} 
\subsection{Topology-aware Node Influence}
We define \textit{Routing Subtree} $\mathcal{RT}^S$ for each destination-oriented tree graph $\mathcal{RT}^S =(N^S, L^S,c(t))$, rooted at the LLN sink, $s = \{1,2,3, ... \mathcal{N}\}$ with $\mathcal{RT}^S \subseteq \mathcal{G}$, $N^S \subseteq \mathcal{N}, L^S \subseteq \mathcal{L}$. Note that $N^S = \{ N_1, N_2, ..., N_S\}$ is a partition of the set of all sensor nodes $N^S$ and $L^S$ is the set of direct links between each node $i$ to preferred parent $p_i$, so $|L^S|= |N^S|-1$. For each node $n \in \mathcal{N}$, we can identify the subtree size of node $n$, $ST_n(t)$ is composed by all the nodes connected to $n$ through multi-hop paths. 
\begin{equation}
ST_n(t) = |N^S_n|
\end{equation}
We define the notion of \textit{Node Influence} $NI_{n,p}(t)$ to measure the influence of parent candidate $p$ to new joining node $n$ at time $t$. Intuitively, the \textit{Node Influence} is determined by sum of the subtree size of potential preferred parent and ETX value between the new joining node and the parent candidate. 
\begin{equation}
NI_{n,p}(t) = \alpha ST_p(t) + \beta ETX_{n,p}(t)
\end{equation}
As the above equation, \textit{Node Influence} of potential parent $p$ to new joining node $n$ is a combination of subtree size of node $p$, $ST_p(t)$, and link quality between $n$ and $p$, $ETX_{n,p}(t)$. We use two weighted number $\alpha$ and $\beta$ to control the interaction between the skew of DODAG subtrees and path quality, this problem is described in more detail in Section IV.  
\subsection{ RPL Control Message DIO extension in support of balancing routing }
Basically, RPL supports various routing metric such as hop-count, ETX, latency, and energy. In order to construct a DODAG, the root broadcasts Destination Information Object (DIO) control messages to the other nodes in the downward direction. It  plays an important role by helping nodes in discovering RPL Instances with their configuration parameters and in constructing a DODAG. The structure of the DIO message is described in Fig.~\ref{fig:diostructure}.

In order to achieve topology balance, RPL control message DIO\footnote{DIO has 16 reserved bits, we use 8 bits to deliver \textit{Routing Subtree} size information, representing the number of child nodes} is exploited to broadcast \textit{Routing Subtree} size information $ST_p(t)$ of a node to its own neighbors. SB-RPL makes all nodes transmit amended RPL DIO messages to its neighboring nodes. When a node $k$ receives a DIO message from a neighbor $n_k$, it records \textit{Routing Subtree} size information $ST_n(t)$ in its neighbor table and uses this value of the received DIO message as a metric to build a multi-hop topology that is free from load imbalance problem.

SB-RPL uses the Trickle Timer to control the broadcasting rate of DIO messages as well. Each node needs to broadcast its updated DIO messages to its one-hop neighbors. In this case, global repair mechanism is not required to perform. Generally, SB-RPL is a fully distributed routing protocol as RPL.     
\subsection{SB-RPL Design}
\begin{algorithm}[t!]
	\caption{: SB-RPL Algorithm}
	\begin{algorithmic}[1]
		\REQUIRE Received DIO messages
		\ENSURE The balance among DODAG subtrees	
		
		\STATE \textbf{Calculation} 
		\STATE \textit{n $\leftarrow$ DIO}; \textit{/*Node n receives a DIO message*/ }
		\IF { n $\notin$ $P_n$}
		\STATE $P_n$ $\rightarrow$ n
		\ENDIF
		
		\IF {\textit{ n == $P'_n$}}
		\STATE $C_n$ $\leftarrow$ $C_n$ $+$ 1 % \textit{Child$\_$Set $\leftarrow$ Child$\_$Set + 1}
		\ENDIF \\
		\COMMENT {\textit{/*
				Rank Computation based on "Node Influence" factor; \\
				$NI_{n,p}(t)$ = $\alpha ST_p(t)$ $+$ $\beta ETX_{n,p}(t)$; \\
				$ST_p$ and $ETX_{n,p}$ are smoothed by EWMA filter. */}}
		
		\STATE $\mathcal{R}_n$  = $\mathcal{R}_{n,p}$ + $NI_{n,p}(t)$\\
		
		\COMMENT{\textit{/* Parent Selection Procedure */}}
		\IF { $R_n$ $>$ $R_{received DIO}$}
		\STATE Maintain the location of node n in the DODAG
		\STATE \textbf{break};
		\ELSE
		\STATE \textit{/*Get the parent with lower rank \\ Discard the current rank*/}			
		
		\STATE $\mathcal{R}_{n,p1} \leftarrow$ \textit{parent$\_$path$\_$metric(p1)}
		\STATE $\mathcal{R}_{n,p2}\leftarrow$ \textit{parent$\_$path$\_$metric(p2)}
		
		\IF {\textit{p1==current$\_$parent $||$ p2 == current$\_$parent}}
		\IF {$\mathcal{R}_{n,p1}$ $<$ $\mathcal{R}_{n,p2}$ $+$ $\sigma$  $\&\&$  $\mathcal{R}_{n,p1}$ $>$ $\mathcal{R}_{n,p2}$ - $\sigma$}
		\STATE \textbf{return} $P'$ \textit{ /*Preferred parent*/}
		\ENDIF
		\ENDIF	
		\ENDIF
		\STATE \textbf{Broadcast} the updated DIO messages.			
	\end{algorithmic}
\end{algorithm}
At the beginning, the DODAG sink propagates DIO messages periodically to its neighbors, with the information mentioned in section II. After a node receives this DIO message, it will decide whether it will join the DODAG or not and computes its own rank once it joins DODAG. 

In this part, we describe the detail of proposed scheme SB-RPL. In order to optimize the balanced of DODAG routing topology, we exploit the new metric, \textit{Node Influence} $NI_{n,p}(t)$ which is introduced above (Equation 10). \textit{Node Influence} $NI_{n,p}(t)$ is considered as a new constraint in the DAG Metric Container included in the DIO control message. In which, the subtree size information is included, and SB-RPL use the $ST_p(t)$ to avoid traffic congestion as well as balance the DODAG subtrees. 
For updating ETX and \textit{Subtree Size}, SB-RPL uses the same EWMA filter as in standard RPL. 
SB-RPL uses a hysteresis mechanism similar to the one employed in MRHOF to prevent unstable changes during fast fluctuations in routing.

Then, each node generates a parent candidate set from its own neighbors according to equation 5, and selects the best parent node according to equation 11. For SB-RPL, SB-RPL nodes compute rank as followed:
\begin{equation}
\begin{split}
\mathcal{R}_n(t) & = \mathcal{R}_{p}(t) + NI_{n,p}(t)  \\
 & = \mathcal{R}_{p}(t) + \alpha ST_p(t) + \beta ETX_{n,p}(t)
\end{split}
\end{equation}

In standard RPL, the DIO transmission procedure based on \textit{Trickle Timer} which is reset to a minimum when there are changes in routing topology. Being a preferred parent of many children results in more traffic overhead and imbalanced problem, consequently consuming its own power much faster than other candidate parents. In order to figure out the problem, we exploit the vital information of DODAG subtree sizes as well as consider the quality of the transmission medium. In other words, the parent with the fewer number of children will be priority selected as the preferred parent. The skewness and balancing of RPL are achieved by reducing the number of children within each subtree of the overloaded bottleneck node. Consequently, joining node will prefer choosing parent according to the dedicated routing metric and guaranteed that the preferred parent has less number of children, equivalent to less size of the subtree. Besides, SB-RPL also considers the link characteristic in the routing procedure. Therefore, SB-RPL guarantees the balance of DODAG trees as well as enhances the reliability. The SB-RPL protocol is described in Algorithm 1. %\\

\section{Performance Evaluation}
In this section, we perform an extensive experimental evaluation of the proposed scheme. We compare our proposed scheme to the state-of-the-art, including Objective Function Zero, MRHOF in various scenarios with the numerical simulation environment and real-world platform. %

\subsection{Methodology}
We evaluate the  performance of compared routing strategy by employing both testbed experiments and simulations. 

\subsubsection{Testbed Experiments}
\begin{figure}[t!]
	\centering
	\includegraphics[width=0.95\linewidth]{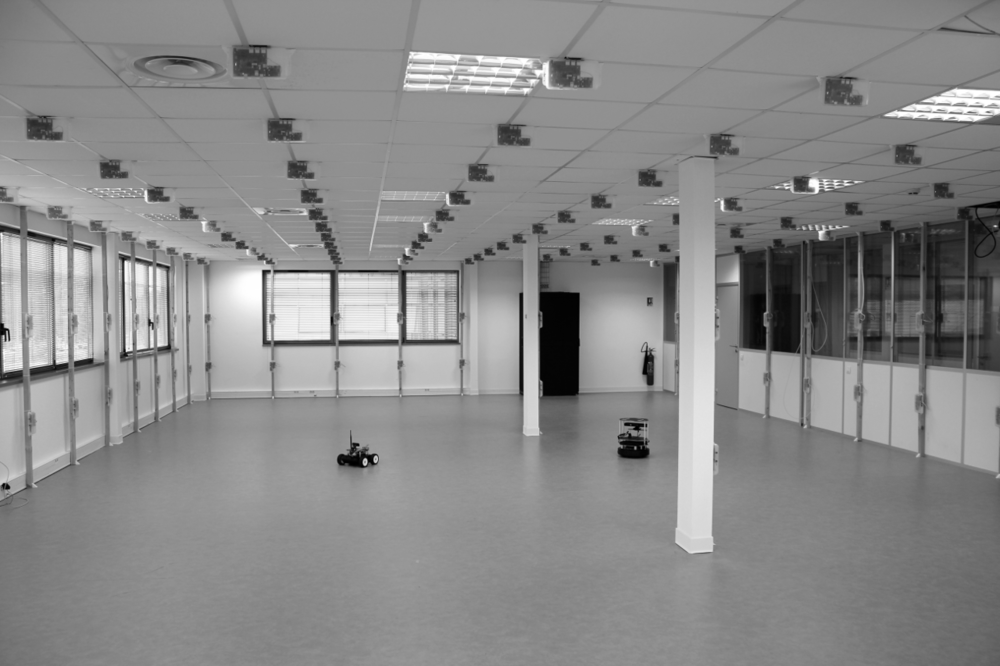}
	\caption[]{\textbf{FIT IoT-Lab Lille side}.Node deployment}.
	\label{fig:lilleIoTLab}
\end{figure}

Setting up a complete WSN deployment is a very complex task. In this part, we present  our study on the FIT IoT-LAB testbed. In our experiments, we used the platform installed in Lille site, France. We used 100 nodes (M3 ARM-Cortex) from Lille site offered by the FIT IoT-Lab testbed, shown in Fig.~\ref{fig:lilleIoTLab}. The topology includes 1 sink located at the center and 99 random sensor nodes generating UDP packets on a predefined time interval. The M3 node has one ARM M3-Cortex micro-controller, one 64kB RAM, one IEEE 802.15.4 radio AT86RF231, one rechargeable 3.7V LiPo Battery and several types of sensors. In order to ensure multi-hop topology is constructed, we set transmission power to -17dBm as in the tutorial of FIT IoT-Lab testbed. The detail of parameters is described in Table II. %

\subsubsection{Cooja Simulation}
To compare our proposed scheme against prior works with full control in network conditions, we use Cooja - the network simulator of Contiki. Cooja provides three different radio models namely UDGM(Unit Disc Graph Mode) - distance loss, UDGM(Unit Disc Graph Model)- constant loss, Multi-Path Ray-Tracer Medium (MRM). In this paper, we use the MRM model because MRM is the most realistic model for wireless sensor network implementation. MRM considers concepts such as reflection, refraction, diffraction, and fading\cite{cooja}. Cooja emulates nodes running compiled MSP430 firmware. Cooja allows us to have completed control over network conditions and emulate varying connectivity. However, as soon as the growth up of network scales, the neighbor tables of nodes start falling apart. Therefore, a straightforward method to solve this issue is to scale up the RAM beyond the expected size of the network. Running simulation with lots of nodes is very CPU consuming. To speed up Cooja simulator, we run all out non-GUI simulations on cloud servers. Each server runs Ubuntu 13.04 LTS with 32 GB of RAM. We evaluate the impact of network scale and density to DODAG topology constructing as well as skewness in various scenarios.%

\subsection{Compared Objective Functions}
We compare our proposed scheme to Objective Function Zero, MRHOF using ETX and MRHOF using $ETX^2$, three state-of-the-art objective functions, which are all implemented in ContikiOS. 
\begin{itemize}
	\item Objective Function Zero (OF0): OF0 uses hop-count as a routing metric. A node calculates its rank by adding a positive and indirectly normalized scalar value to its preferred parent rank. This objective function can also be called as minimum hop-count objective function.
	\item Minimum Rank with Hysteresis Objective Function using ETX (MH$\_$ETX): MRHOF selects routes that minimize additive routing metrics such as energy, latency, and ETX. In addition, MRHOF uses hysteresis to reduce instability due to small metric changes. In ContikiOS, MRHOF is used with ETX routing metric as default objective function. 
	\item Minimum Rank with Hysteresis Objective Function using $ETX^2$ (MH$\_$ETXSQ): An extended version of MRHOF, however, MH$\_$ETXSQ uses $ETX^2$ as the routing metric.   
\end{itemize}
The source code of objective functions is fairly provided on ContikiOS homepage\cite{contikisource}.% 

\subsection{Metrics}
We focus on 3 types of metrics to compare the performance of the objective functions: skewness metrics, reliability and latency. 
\subsubsection{Skewness Indexes}
In order to compare the skewness of our proposed routing scheme to existing objective functions of RPL standard(OF0, MRHOF), we define 4 skewness indexes namely $\mathcal{M}1$, $\mathcal{M}2$, $\mathcal{M}3$, and $\mathcal{M}4$ for each rank of DODAG. The skewness indexes determine the extension of asymmetry or lack of symmetry among subtrees of a DODAG. The definitions of skewness indexes are defined as follows: 

$$
\mathcal{M}1 = \frac{ST^{\max}(t) - ST^{\min}(t)}{ST^{avr}(t)}; \quad\quad\quad\quad
\mathcal{M}3 = \frac{ST^{\max}(t)}{ST^{\min}(t)} \\
$$
$$
\mathcal{M}2 = \frac{ \sum_{i=1}{\left|ST_i(t) - ST^{avr}(t) \right|}}{ST^{avr}(t)};\quad
\mathcal{M}4 = \frac{ST^{\max}(t) - ST^{\min}(t)}{ST^{\min}(t)} \\
$$

where $ST^{max}(t)$, $ST^{min}(t)$, and $ST^{avr}(t)$ are the maximum value, minimum value and average value of the subtree sizes in a DODAG, respectively. $\mathcal{M}1$, $\mathcal{M}2$, $\mathcal{M}3$, and $\mathcal{M}4$ are used to measure the skew and balancing of DODAG. The minimum value of skewness indexes are 0 when the size of subtrees are ideally equal, and a smaller $\mathcal{M}1$, $\mathcal{M}2$, $\mathcal{M}3$ and $\mathcal{M}4$ values indicate better balancing performance. For example, in Fig. 1, DODAG includes 1 RPL sink and 10 source nodes. Nodes $\{a,b,c\}$ are level 1, nodes $\{d,e,f,g\}$ are level 2 and nodes $\{h,i\}$ are level 3. For level 1 at time $t$, $ST^{max} =ST_{(a)} = 3$, $ST_{min} = ST_{(b)} = ST_{(c)}= 2$, $ST^{avr} = 2.34$, so $\mathcal{M}1=0.43$ , $\mathcal{M}2=0.86$, $\mathcal{M}3=1.5$ and $\mathcal{M}4=0.5$. The skewness value of higher  is computed similarly. In this paper, we compute the skewness values of three levels $\{1,2,3\}$ and aim to minimize the value of skewness indexes.
In this paper, we aim to minimize the four skewness indexes subject to number of nodes in each DODAG subtree. In our experiments, we also use two specify metrics namely Packet Delivery Ratio and Latency to evaluate the performance of SB-RPL. 
%\begin{equation*}
%\begin{aligned}
%& \text{minimize} 
%& & \mathcal{M}1;\mathcal{M}2;\mathcal{M}3;\mathcal{M}4; \\
%& \text{subject to}
%& &\mathcal{R}_{p}(t) \\
%%&&&  NI_{n,p}(t).
%\end{aligned}
%\end{equation*}

\begin{table}[t!]
	\caption{FIT-IoT-Lab Experimental Setup}
	\label{table2}
	\begin{tabular}{p{4cm}  p{4cm}}
		\hline
		\hline
		\textbf{Experimental Parameters}  & \textbf{Values}  \\ 
		\hline
		\hline
		Environment & Indoor \\	
		Network Scale & 99 nodes and 1 sink (center) \\  
		Node spacement &  uniform random\\
		Deployed nodes & 100 random nodes \\
		Platform & ContikiOS/M3 Cortex ARM \\
		Duration & 60 min per instance \\
		Application Traffic & UDP/IPv6 traffic \\
		Payload size & 16 bytes \\
		Number of hops & Multihop \\
		Embedded network stack & ContikiMAC \\
		Number of Retransmissions & 10 Retransmissions \\
		Compared Objective Functions &  RPL (OF0, MRHOF), SB-RPL \\
		
		\hline
		\hline
		\textbf{Hardware Parameters} & \textbf{Values} \\
		\hline
		\hline
		Antenna Model & Omni-directional \\
		MAC & 802.15.4 beacon enabled \\
		Radio Chip & TI CC2420 \\
		Radio propagation & 2.4 GHz \\
		Transmission Power & -17 dBm \\
		%		Radio Channel & Channel 26 \\
		RX RSSI threshold & -69 dBm \\
		\hline	    
	\end{tabular}
\end{table}

\subsubsection{Packet Delivery Ratio}(PDR) is the ratio of the number of packets that are successfully delivered to a destination over the number of packets that are sent by the transmitter in an end-to-end communication. PDR represents the reliability of the routing protocol. In most cases, PDR is the important evaluation metric of a network.
\begin{equation*}
\begin{aligned}
\text{Average PDR} = \frac{\text{Total Packets Received}}{\text{Total Packets Sent}}* 100
\end{aligned}
\end{equation*} 
\subsubsection{Average Latency} represents the end-to-end latency on the application. Latency is the time elapsed from the application on the source node handling the packet to the MAC layer until the packet arrives at the sink's collection application.  Minimizing latency is one of the main targets of routing protocol design. %
\begin{equation*}
\begin{aligned}
\text{Average Latency} = \frac{\sum_{k=1}^{m}(\text{RecvTime(k)-SentTime(k)}} {\text{Total Packet Received}}
\end{aligned}
\end{equation*}
where \textit{m} is the total number of packet received successfully. In the simulations, we use the timing information provided by Cooja Simulator.

\subsection{Testbed Experiments}
The extensive experiments were conducted to compare practical performance of the proposed scheme to existing RPL routing standards based on random 100-node topologies in the FIT IoT-LAB tested.

\begin{figure}[t!]
	\centering
	\includegraphics[width=0.95\linewidth]{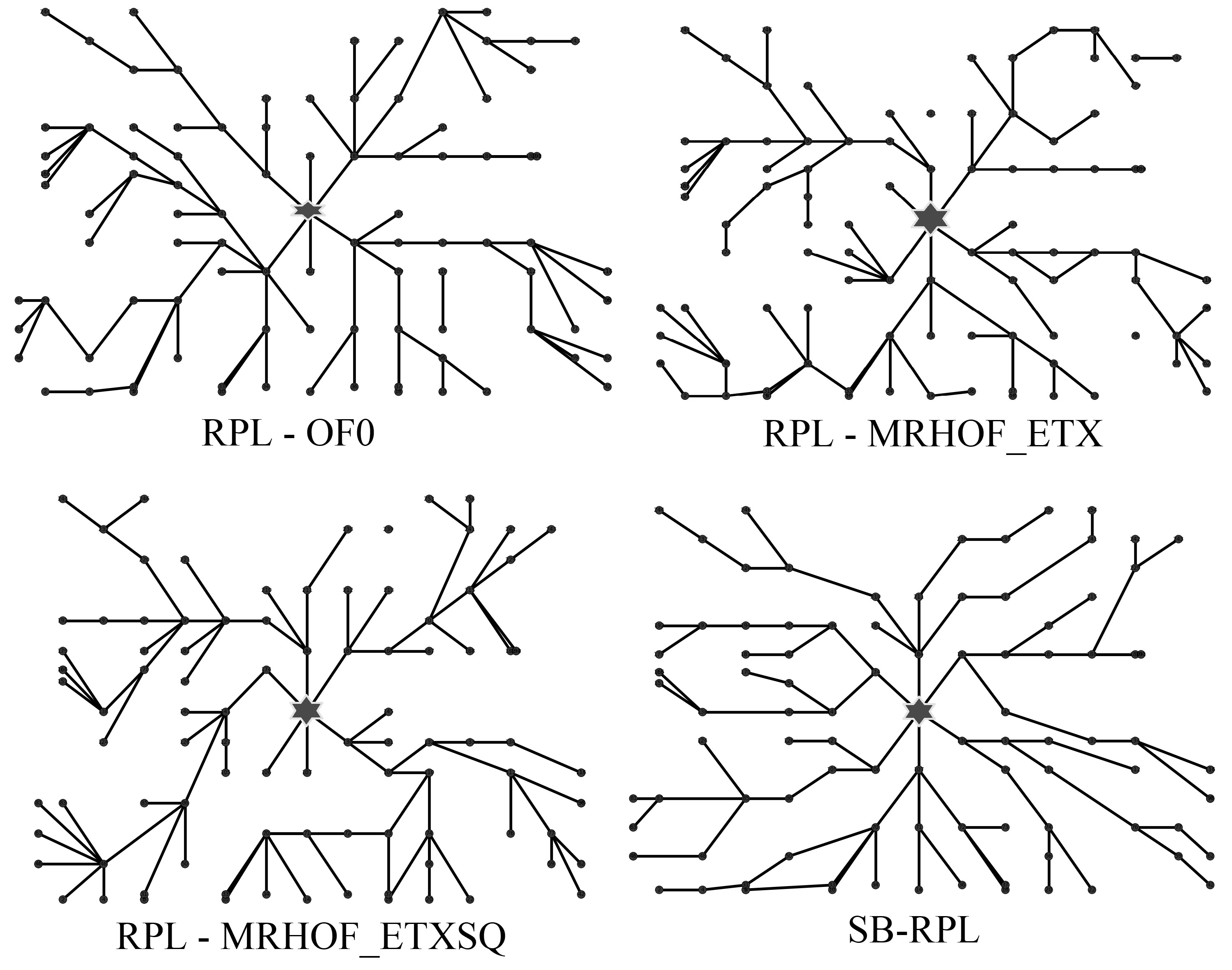}
	\caption[]{\textbf{RPL DODAG Routing Topologies}. Snapshots of routing topologies of Objective Functions implemented on FIT-IoT-Lab platform. OF0 and MRHOF show an unbalance among subtrees in DODAG because they use simple routing metrics for parent selection procedure, while SB-RPL not only considers the reliability of data transmission but also the skew and balance of DODAG.}
	\label{fig:TopoComparison}
\end{figure}%

\subsubsection{Impact of $\alpha$ and $\beta$}
\begin{figure}[t!]
	\centering
	\subcaptionbox{Reliable }{\includegraphics[width=0.9\linewidth, height=0.12\textheight]{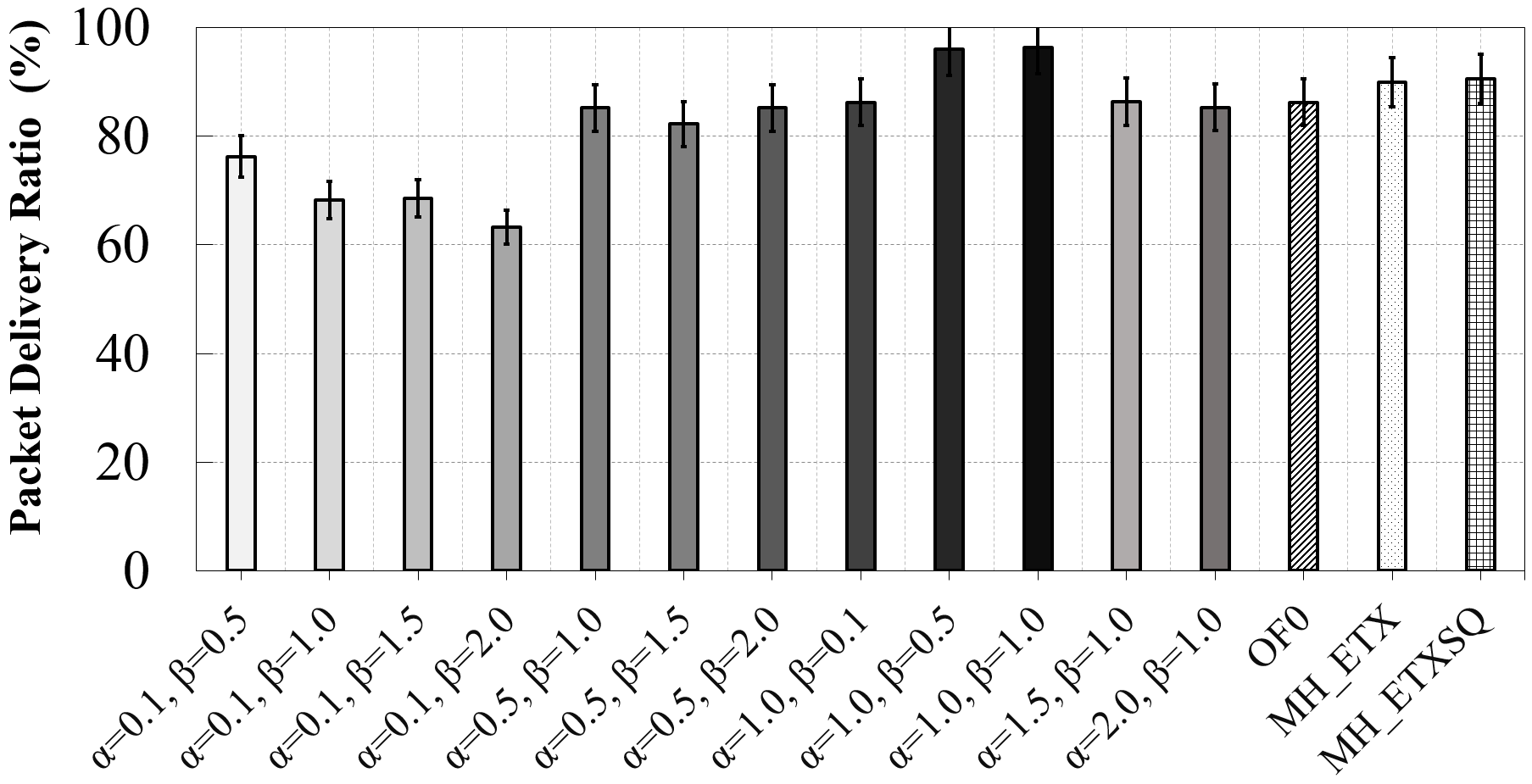}}
	\subcaptionbox{Average Latency (s)}{\includegraphics[width=0.9\linewidth, height=0.12\textheight]{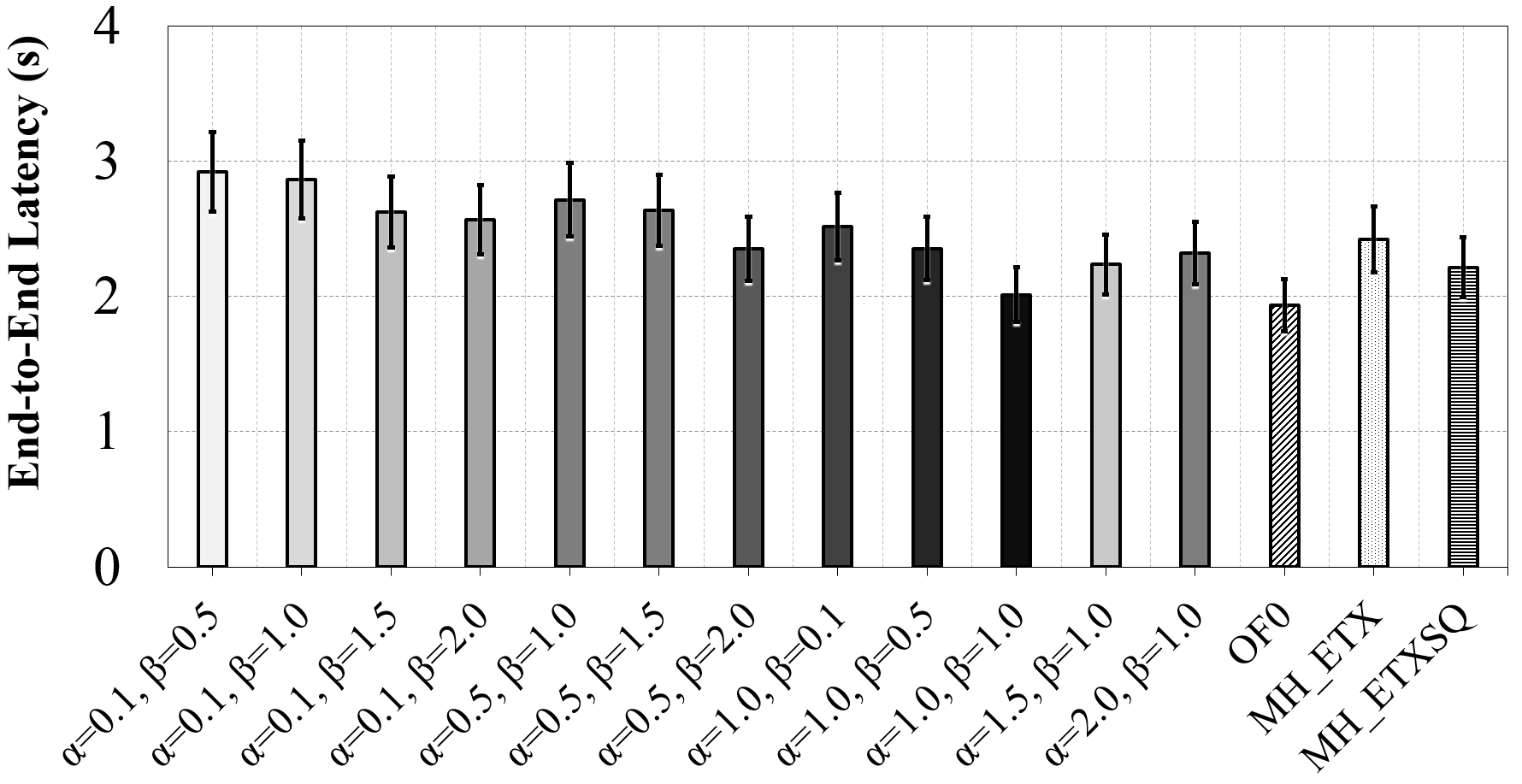}}
	\caption{\textbf{ Impact of $\alpha$ and $\beta$.} The average reliability and latency of SB-RPL are affected by the varying of $\alpha$ and $\beta$.}
	\label{fig:impactpdrdelay}
\end{figure}

\begin{figure*}[t!]
	\centering
	\subcaptionbox{Average $\mathcal{M}1$ metric}{\includegraphics[width=0.23\linewidth, height=0.13\textheight]{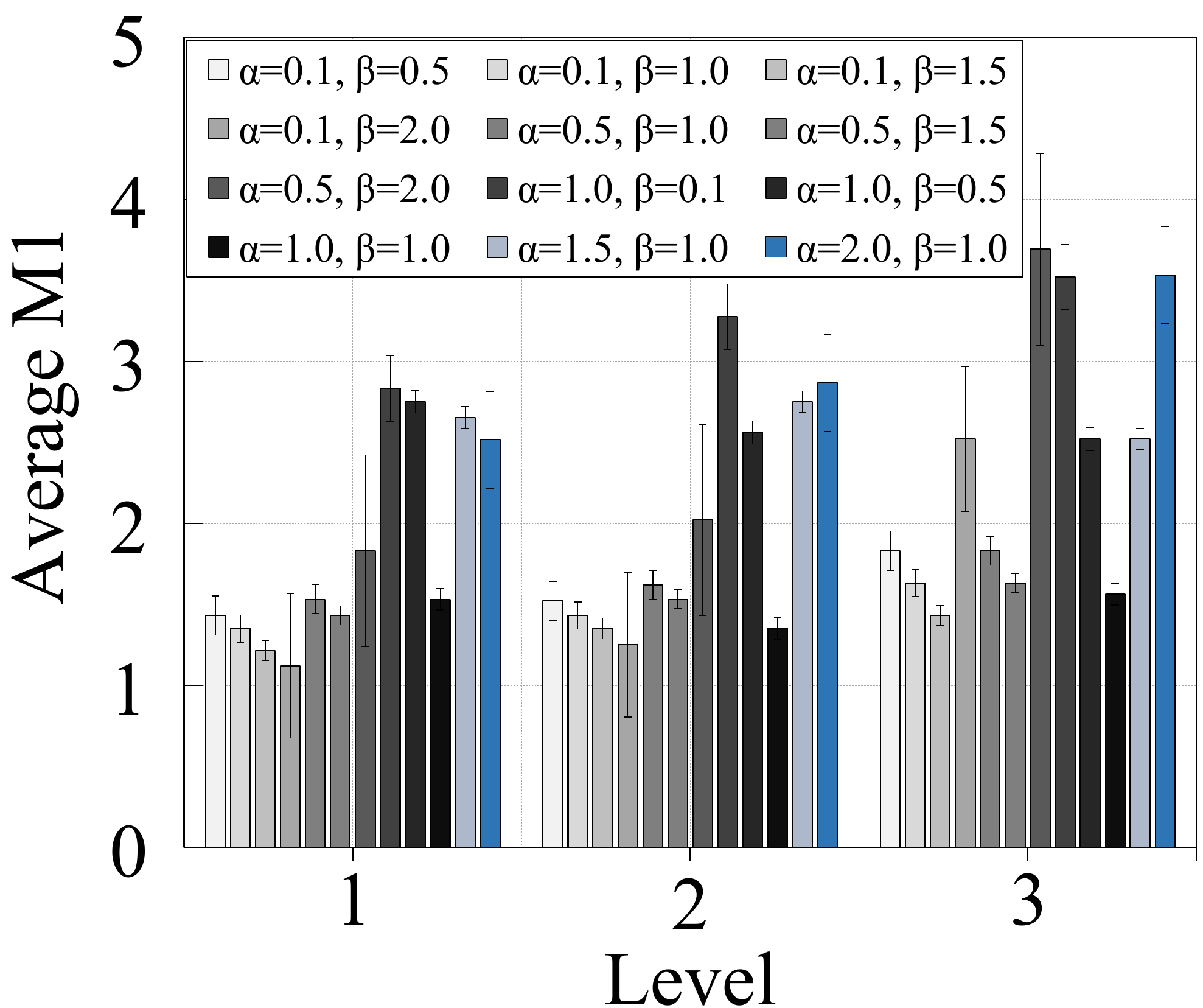}} 
	\subcaptionbox{Average $\mathcal{M}2$ metric}{\includegraphics[width=0.23\linewidth, height=0.13\textheight]{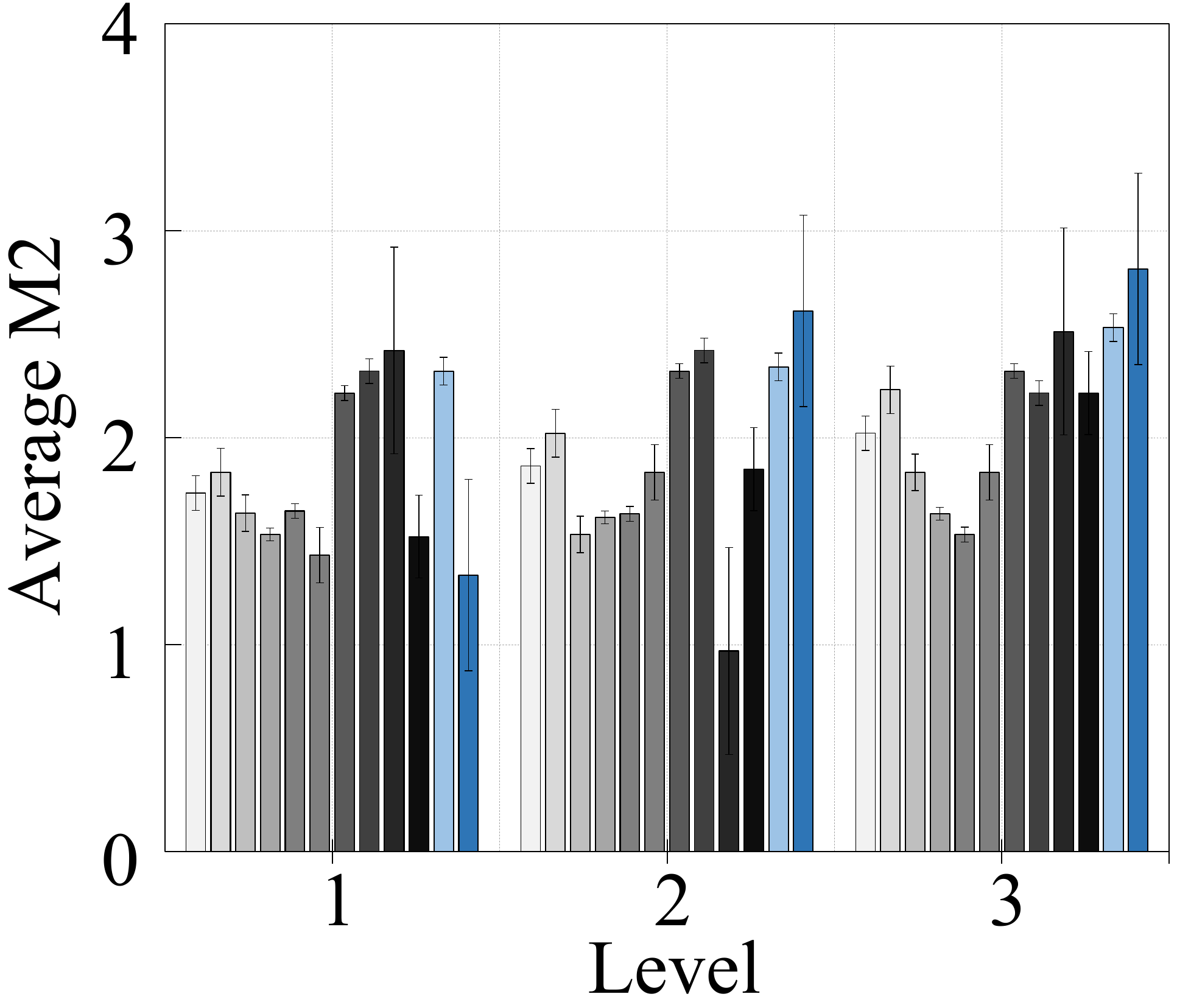}}
	\subcaptionbox{Average $\mathcal{M}3$ metric}{\includegraphics[width=0.23\linewidth, height=0.13\textheight]{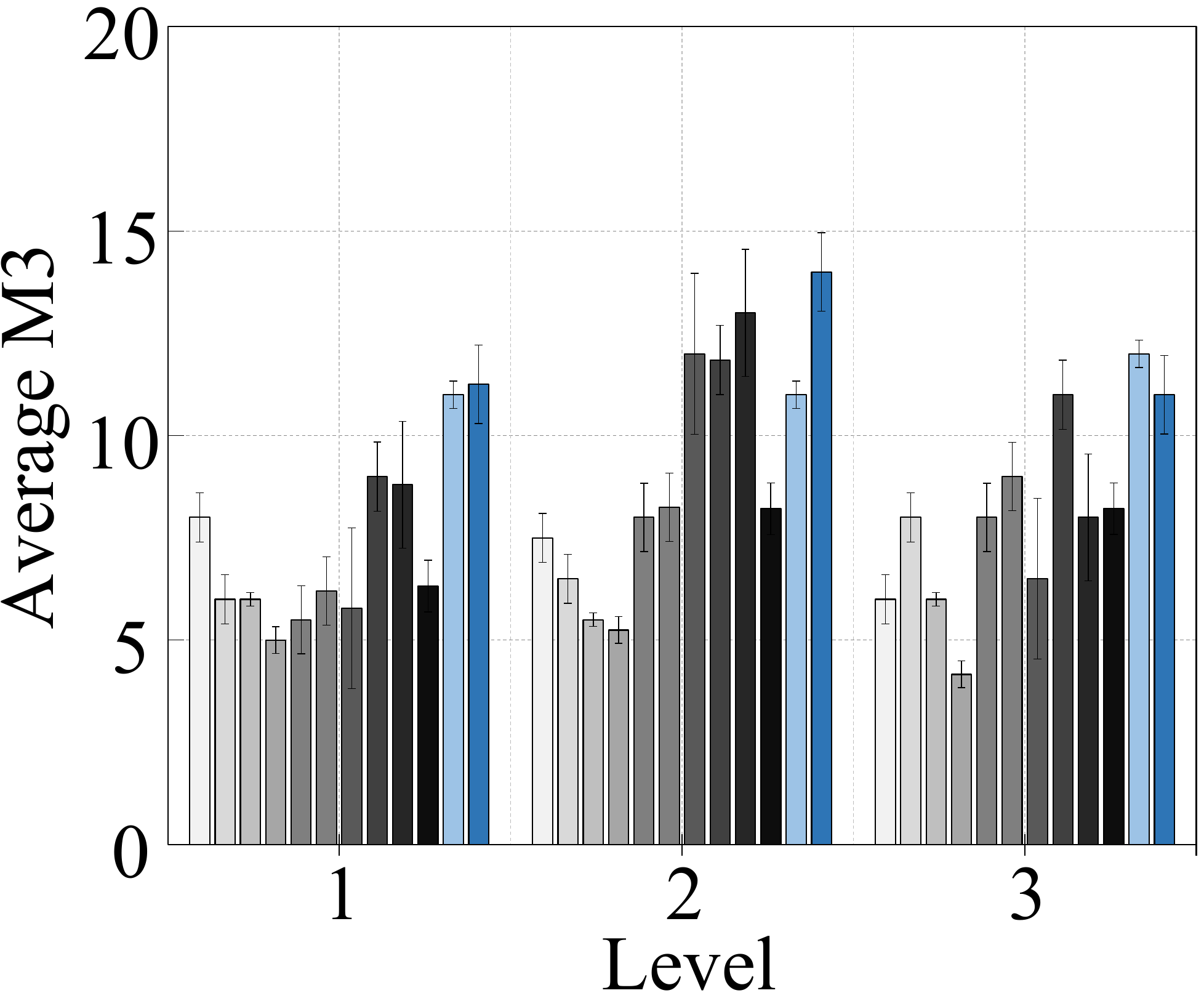}}
	\subcaptionbox{Average $\mathcal{M}4$ metric}{\includegraphics[width=0.23\linewidth, height=0.13\textheight]{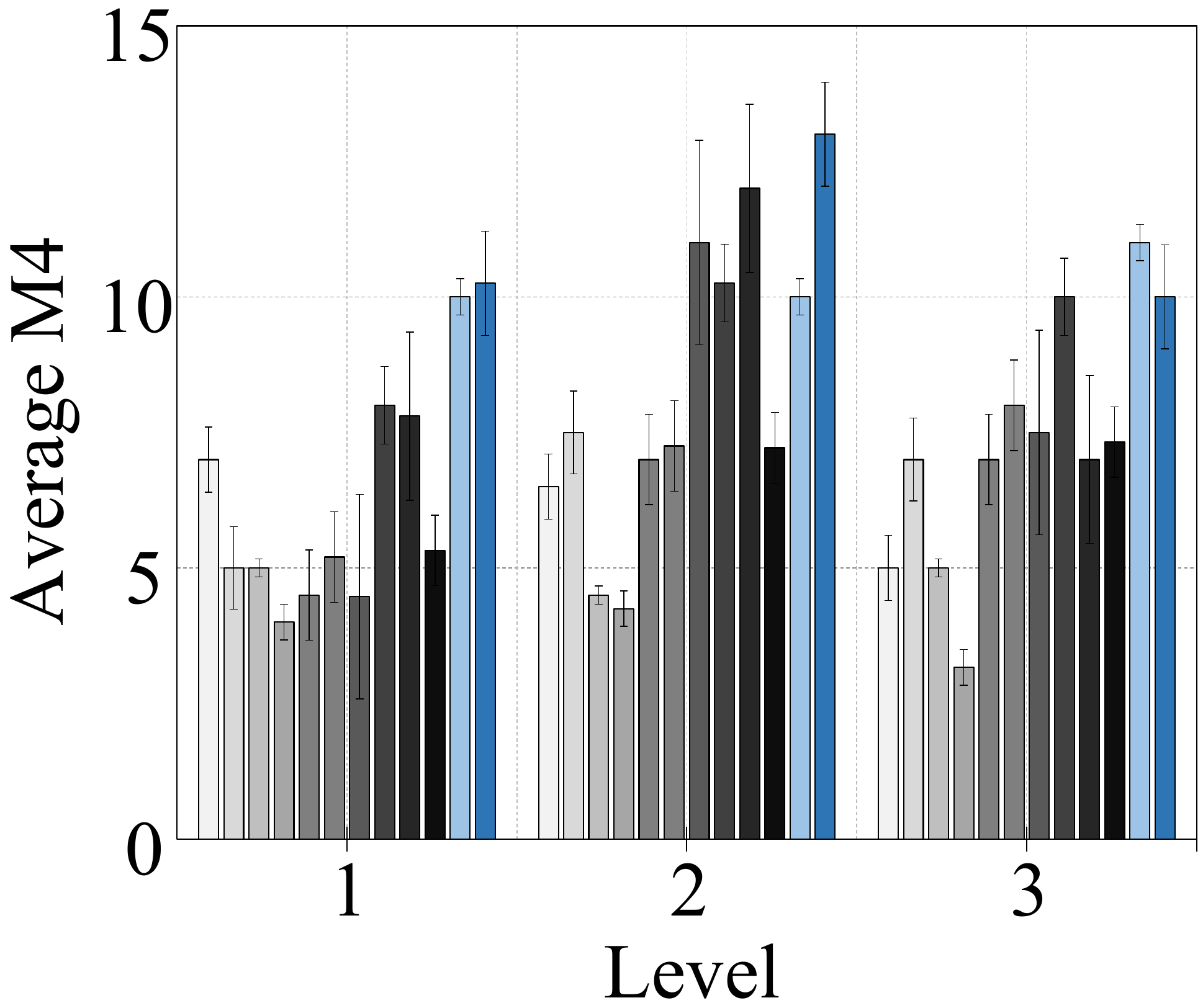}}
	\caption{\textbf{ Impact of $\alpha$ and $\beta$}. We evaluate the skewness and balancing of RPL by measuring four skewness indexes with the various pair of values of $\alpha$ and $\beta$ through FIT-IoT-Lab 100-node topologies. The skewness and balancing of SB-RPL are different in three levels of nodes in the DODAG. From left to right, the order of pair values of $\alpha$ and $\beta$ legend is similar to Fig.~\ref{fig:impactpdrdelay}.}
	\label{fig:abSkew}
\end{figure*}

We investigate the impact of the design parameters $\alpha$ and $\beta$ values on the performance of SB-RPL. Through extensive experiments with different values of $\alpha$ and $\beta$ in a range from $0.1$ to $2$ (Fig.~\ref{fig:impactpdrdelay} and Fig.~\ref{fig:abSkew}), it shows the trade-off between link quality from new devices willing to join DODAG to  parent candidates and the balance of the subtrees. First of all, Fig.~\ref{fig:impactpdrdelay} shows that the PDR first increases with $\beta$, however the PDR also depends on the value of $\alpha$. This is due to the trade-off between the routing direction and congestion control. For a large value of $\beta$, a node would mainly consider link quality ETX to potential preferred parent when selecting the best link quality. This leads to a parent node have to handle many child nodes and the length of paths from the source nodes to the sink might be stressed through many links. Thus, traffic congestion easily occurs. However, a node may select a path that is longer than the shortest path by considering which parent candidate node has the smallest number of children in the routing table, and connect to that parent node to avoid traffic congestion. This shows the trade-off between load balancing and link quality in the routing procedure.%

The FIT-IoTLab platform allows writing the printouts of each device to a log along with a corresponding timestamp. The timestamp of the time the log is written and totally ordered. Therefore, this introduces some latency regarding when the log is written to file. Fig.~\ref{fig:impactpdrdelay}(b) demonstrates that the end-to-end latency of SB-RPL is impacted by the varying of $\alpha$ and $\beta$, SB-RPL reduces the end-to-end delay with respect to original RPL. Thanks to \textit{Node Influence}, the node not only considers the "good enough" link quality path but also refers to the number of connections to the parent candidate for parent selection procedure. Thus, the nodes have fewer children to manage and packet transmission is easy performed without extra waiting time. 

We believe that both parameters $\alpha$ and $\beta$ do significant impact on the performance of SB-RPL, and they can be optimized empirically by regarding network performance. Fig.~\ref{fig:impactpdrdelay} demonstrates that SB-RPL achieves highest PDR when $\alpha=1.0$ and $\beta=1.0$. Besides, the latency is significant differentiate among pair values of $\alpha$ and $\beta$. In terms of selecting a longer path to avoid traffic congestion, the latency is slightly higher than selecting the shortest path. Therefore, we have exploited these values throughout our practical experiment in actual testbed FIT-IoT Lab as well as simulations and performance evaluation section. %

Fig.~\ref{fig:abSkew} compares the skewness indexes of SB-RPL in terms of varying $\alpha$ and $\beta$ from 0.1 to 2. We evaluate the average skewness indexes in three levels of DODAG. First of all, we observe that the skewness indexes increase as the decrease of $\alpha$, however in order to achieve a good performance, the impact of $\beta$ is required. Because when using the large value of $\alpha$, the skewness indexes also decrease, the SB-RPL mainly focus on exploiting the balanced perspective in the routing procedure, the DODAG tree willing to reach balancing. However, when the ratio between $\alpha ST_p(t)$ and $\beta ETX_{n,p}(t)$ grows up remarkably, the nodes willing to join DODAG only might not select the preferred parent with good link quality enough for data transmission. %

\begin{figure*}[t!]
	\centering
	\subcaptionbox{Average $\mathcal{M}1$ metric}{\includegraphics[width=0.23\linewidth, height=0.1\textheight]{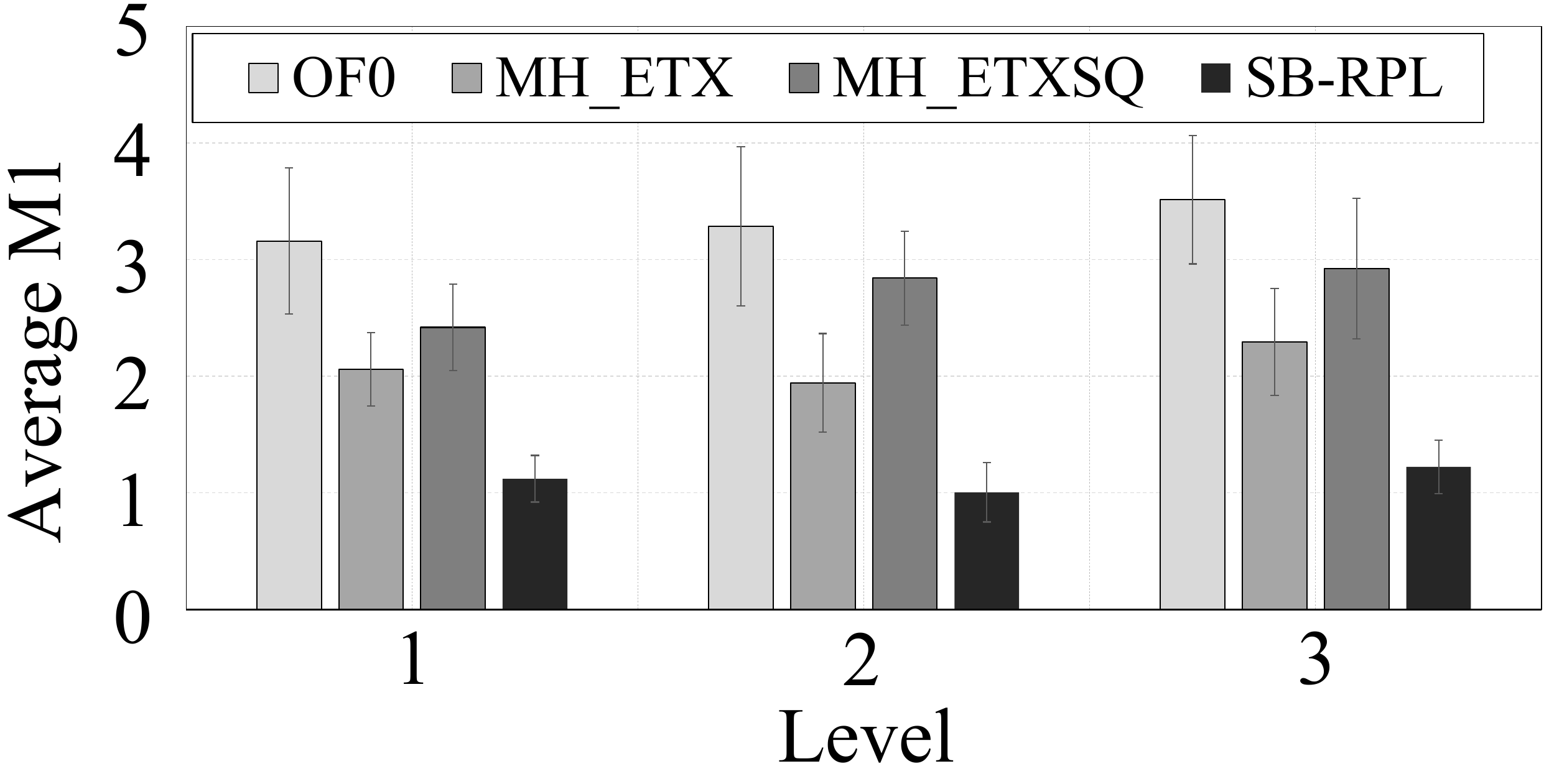}} 
	\subcaptionbox{Average $\mathcal{M}2$ metric}{\includegraphics[width=0.23\linewidth, height=0.1\textheight]{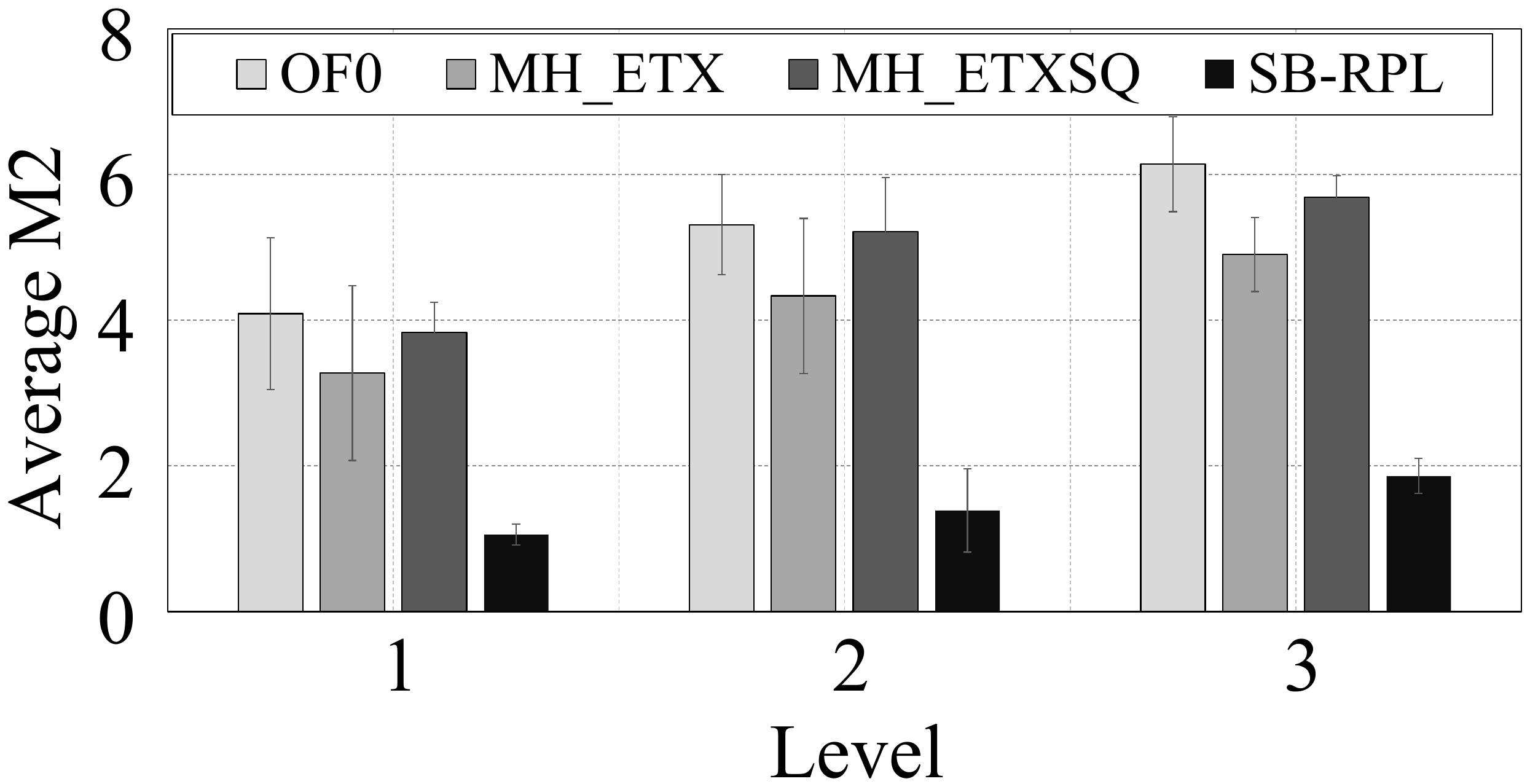}}
	\subcaptionbox{Average $\mathcal{M}3$ metric}{\includegraphics[width=0.23\linewidth, height=0.1\textheight]{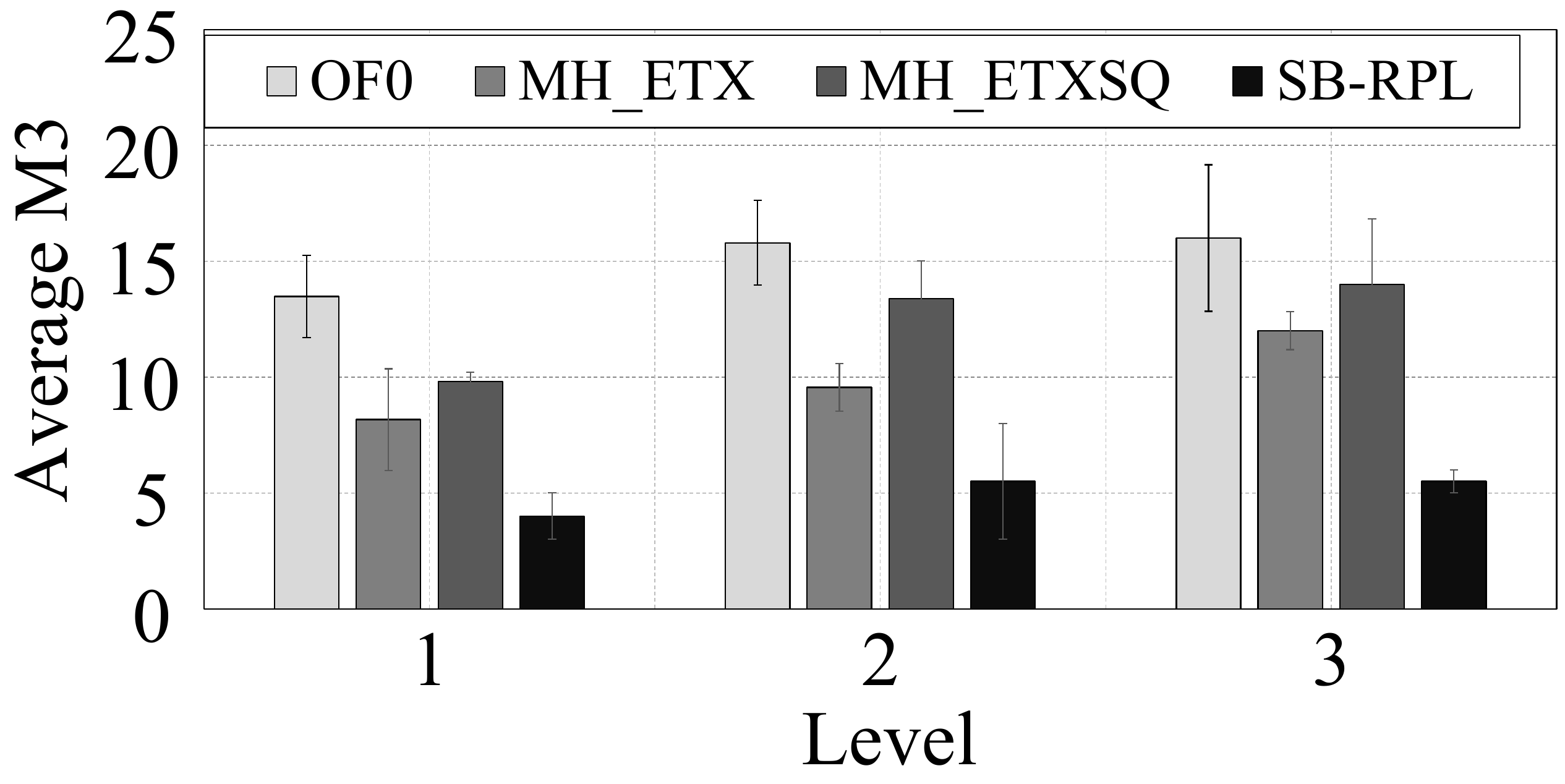}}
	\subcaptionbox{Average $\mathcal{M}4$ metric}{\includegraphics[width=0.23\linewidth, height=0.1\textheight]{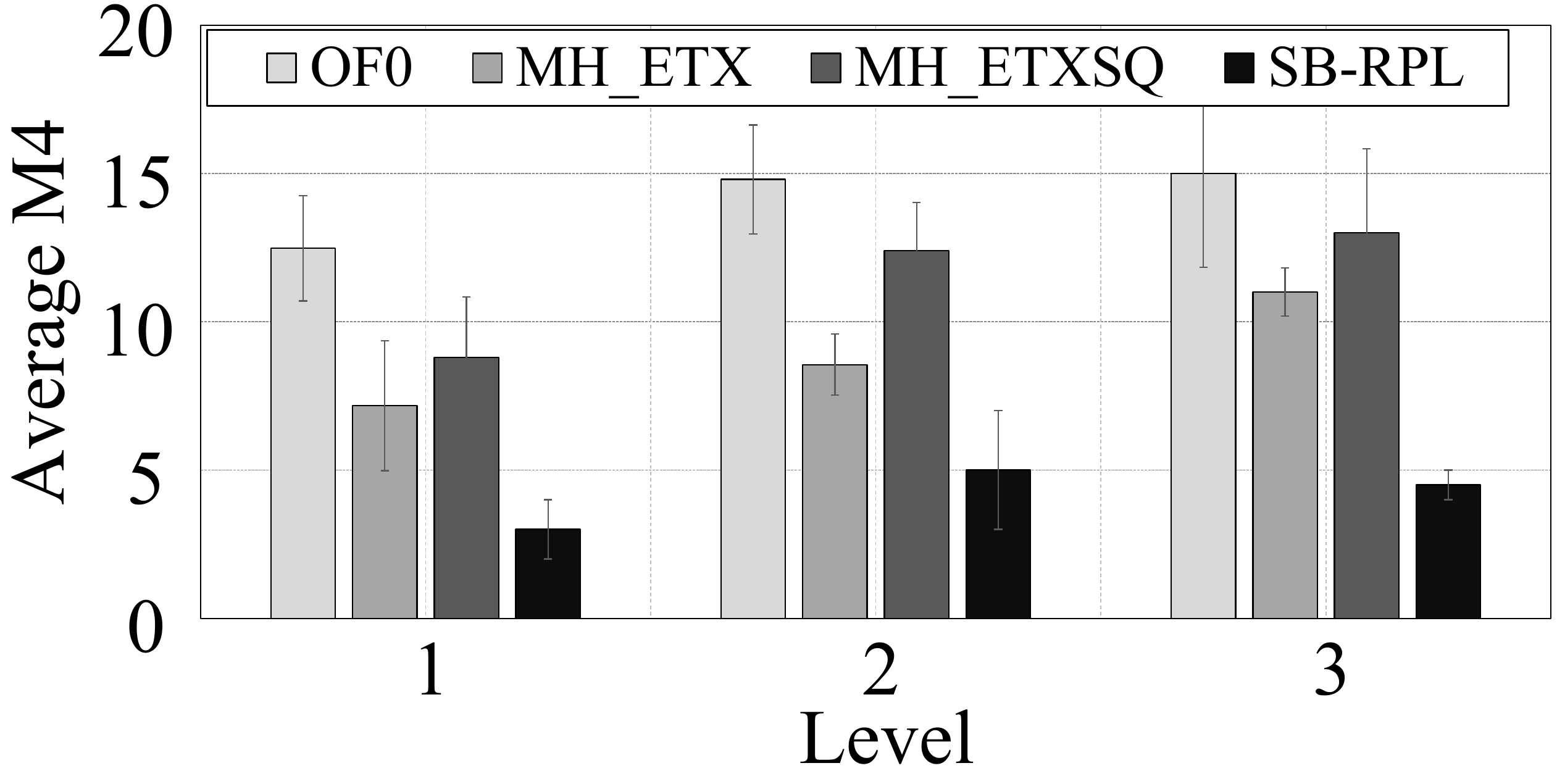}}
	\caption{\textbf{Objective Function Comparison}. We compare the proposed scheme to existing Objective Functions through experiments using the FIT-IoT-Lab platform, Lille site. The skew of SB-RPL is lower around 3 times compared to existing objective functions. Thanks to Subtree size $ST_p(t)$ metric, SB-RPL reaches balancing among subtrees, the parent node does not need to handle too many children. It leads to the number of parent changes reduces significantly. In fact, the child node may converse the same preferred parent for a long time.}
	\label{fig:ofComparison}
\end{figure*}%s

\subsubsection{Objective Function Comparison} We compare our proposed scheme to standard RPL objective functions in terms of skewness and balancing via the practical FIT-IoT-Lab platform. Fig.~\ref{fig:ofComparison} demonstrates that the average skewness indexes of SB-RPL outperform to the rest objective functions in three levels. With 100-node topologies, the average $\mathcal{M}1$ of SB-RPL is around 1 in three levels while OF0, MH$\_$ETX, and MH$\_$ETXSQ are around 3, 2, and 2.5 respectively. Similarly, the skewness indexes $\mathcal{M}2$, $\mathcal{M}3$, and $\mathcal{M}4$ of SB-RPL is less than around 3 times in comparison with other methods. This is because both OF0 and MRHOF use simple routing metrics such as hop count or ETX  for the path calculation and parent selection, meanwhile SB-RPL considers the skew and balance among subtrees and combines multiple metrics for routing efficiency in RPL DODAG. 

\begin{table}[h!]
	\begin{center}
		\caption{Average PDR of compared Objective Functions (Unit: \%)}
		\label{tab:table1}
		\begin{tabular}{|c|c|c|c|} % <-- Alignments: 1st column left, 2nd middle and 3rd right, with vertical lines in between
			\hline
			\cline{2-4} &
			\multicolumn{3}{|c|}{\textbf{Traffic Rates} ($\#$ pkts per second)} \\
			\cline{2-4}
			\hline
			\textbf{Compared Schemes} & 60 \textit{pkts/s} & 40 \textit{pkts/s} & 20 \textit{pkts/s}\\
			\hline
			RPL-OF0 & 62.3\% & 87.1\% & 91.2\%\\
			\hline
			RPL-MH$\_$ETX &  63.5\%  & 88.4\%& 92.2\% \\
			\hline
			RPL-MH$\_$ETXSQ &  65.6\%  & 89.2\%  &92.4\% \\
			\hline
			SB-RPL & 79.6\%  & 94.7\% &96.6\% \\
			\hline
		\end{tabular}	
	\end{center}
\end{table}%
Table III compares the average PDR of routing schemes in three different traffic rates (20 packets per second, 40 packets per second, and 60 packets per second). With SB-RPL, the packet reception rate is enhanced by almost 10\% compared to the original RPL. The loss of packets in real test-bed might be due to the interference from the external world.
Higher the PDR of the network means that the packets lost in the network are less and the link between the nodes are stable. %

\subsubsection{DODAG Routing Topology}Fig.~\ref{fig:TopoComparison} shows an example of routing topologies of compared objective functions. In this experiment, we selected randomly 100 nodes in Lille side of FIT-IoT-Lab platform with the sink was located at the center. 

In cases of standard objective functions OF0, MRHOF, many nodes have selected same one node as their parent and the number of nodes belongs the subtree of this node becomes larger and larger. With OF0, the nodes choose the preferred parent based on the hop-distance to the sink, so the probability of selecting the same preferred parent which has the shortest path to the sink is high. Meanwhile, MRHOF scheme prefers to choose the path with the lowest ETX value, so the length of route from the source node to the sink might be too long. Consequently, the parent node experiences a significant overload and drops a lot of packets transmitted from its child nodes which results significant in reliability degradation.

By contrast, SB-RPL considers the balance among subtrees in the DODAG to achieve balanced traffic load distribution. The new nodes willing to join DODAG obtain the \textit{Subtree Size} information from neighbor nodes and select parent candidate list. From the parent candidate list, nodes compute their own rank by combing multiple factors, in which \textit{Subtree Size} is considered seriously to prevent connecting to a parent node which has too many children. The smart use of SB-RPL results in traffic congestion degradation as well as achieves load balancing of network. %

\begin{figure*}[ht!]
	\centering
	\subcaptionbox{Average $\mathcal{M}1$ - Level 1}{\includegraphics[width=0.24\linewidth, height=0.1\textheight]{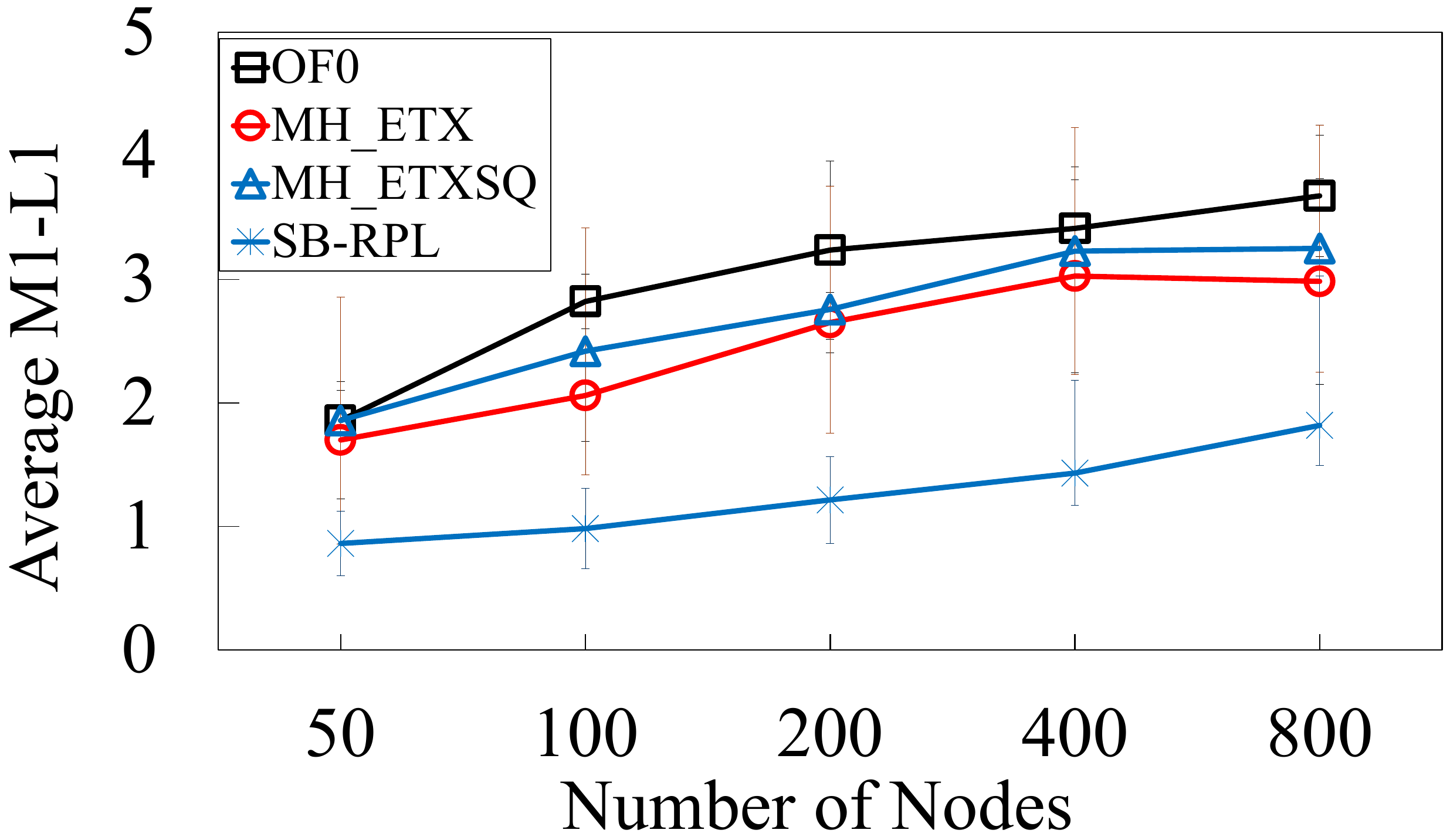}}
	\subcaptionbox{Average $\mathcal{M}2$ - Level 1}{\includegraphics[width=0.24\linewidth, height=0.1\textheight]{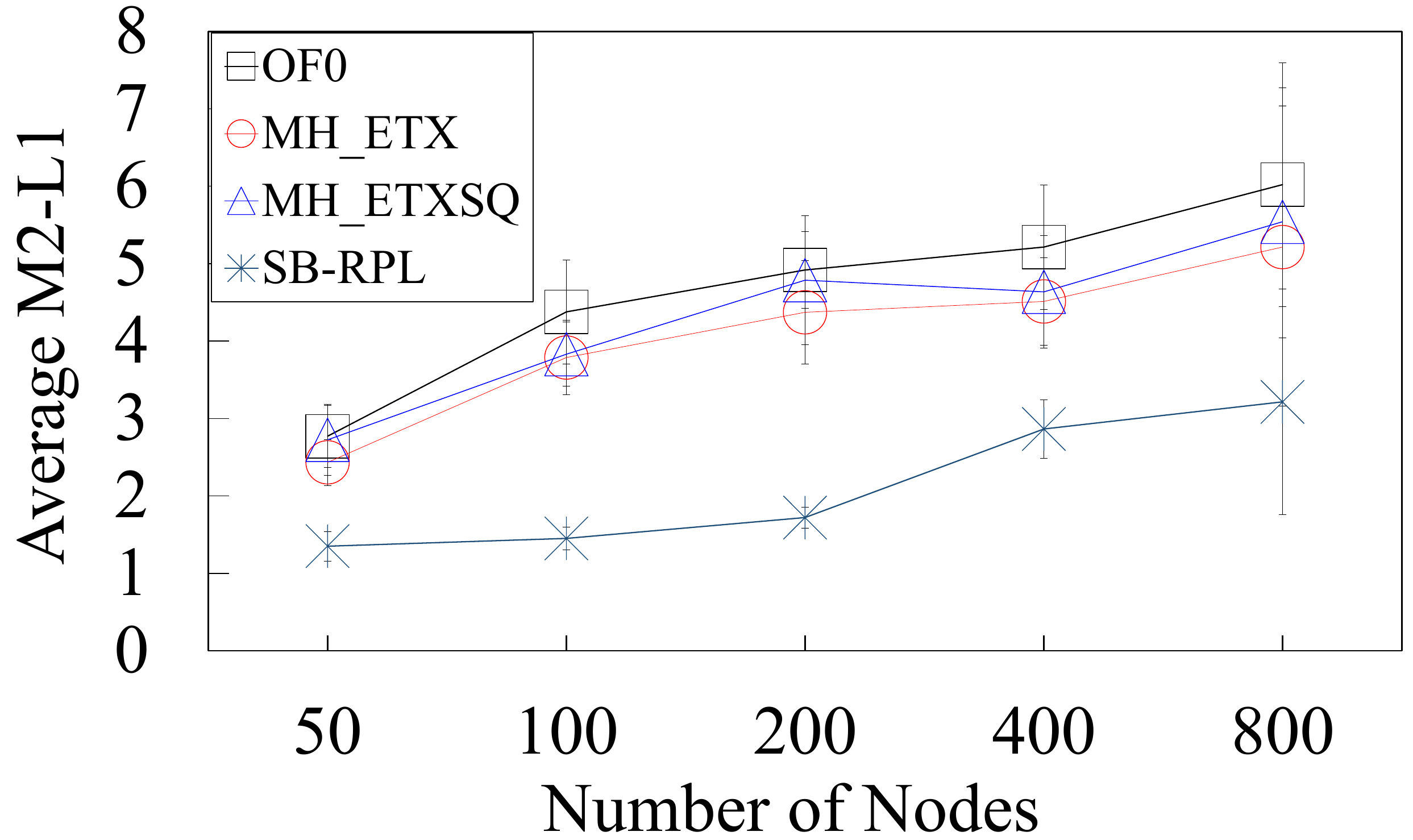}}
	\subcaptionbox{Average $\mathcal{M}3$ - Level 1}{\includegraphics[width=0.24\linewidth, height=0.1\textheight]{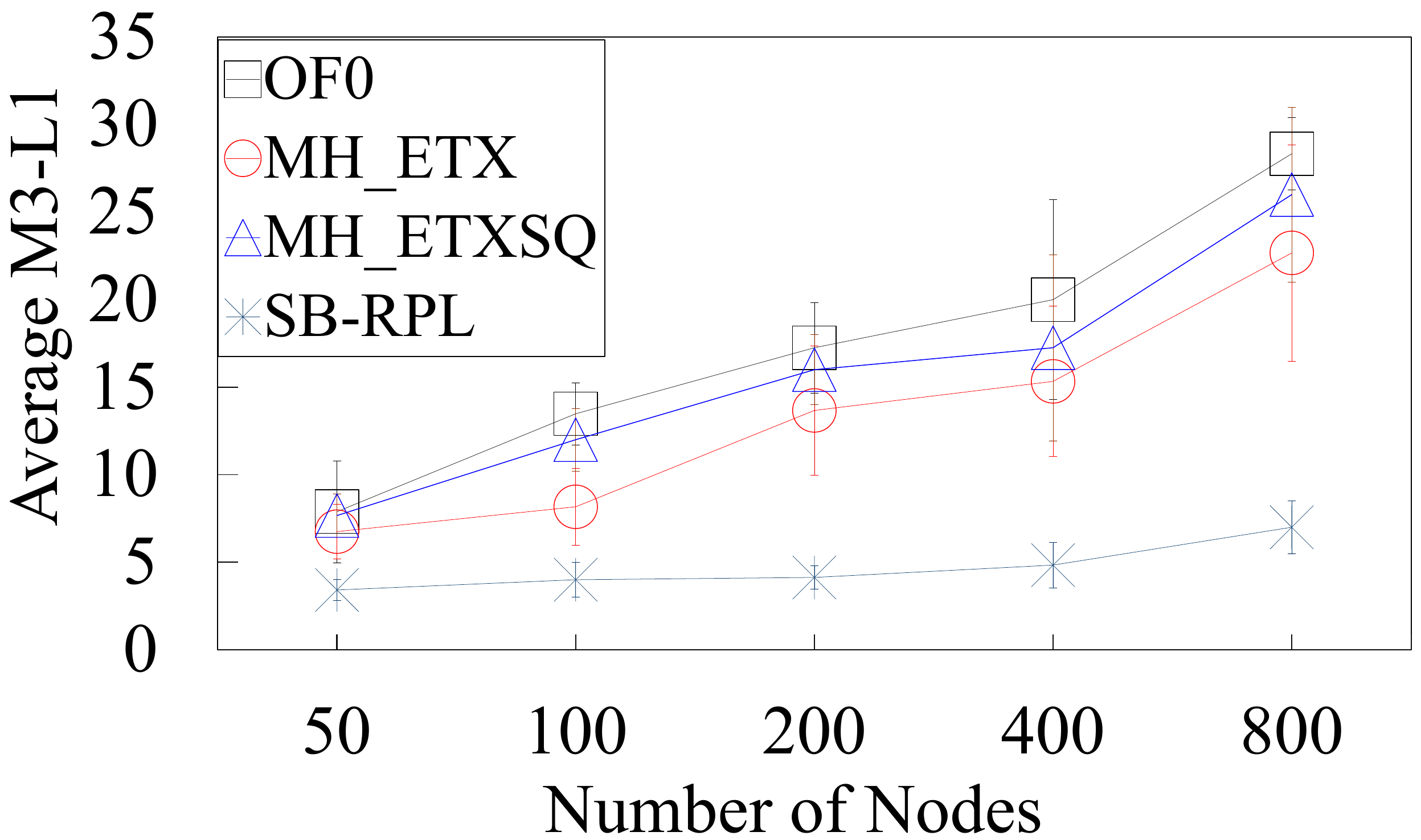}}
	\subcaptionbox{Average $\mathcal{M}4$ - Level 1}{\includegraphics[width=0.24\linewidth, height=0.1\textheight]{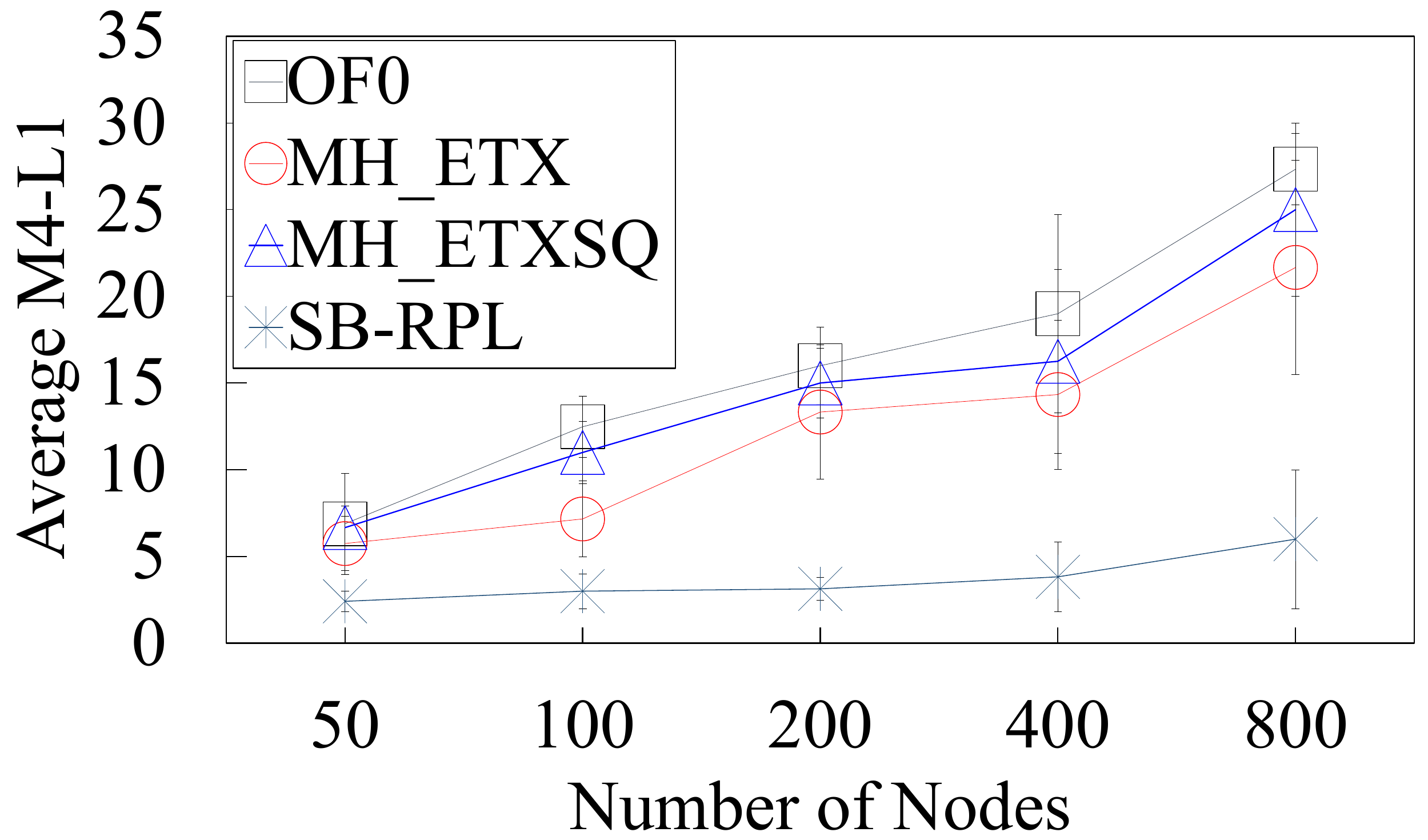}} \\
	\vspace{0.3cm}
	\subcaptionbox{Average $\mathcal{M}1$ - Level 2}{\includegraphics[width=0.24\linewidth, height=0.1\textheight]{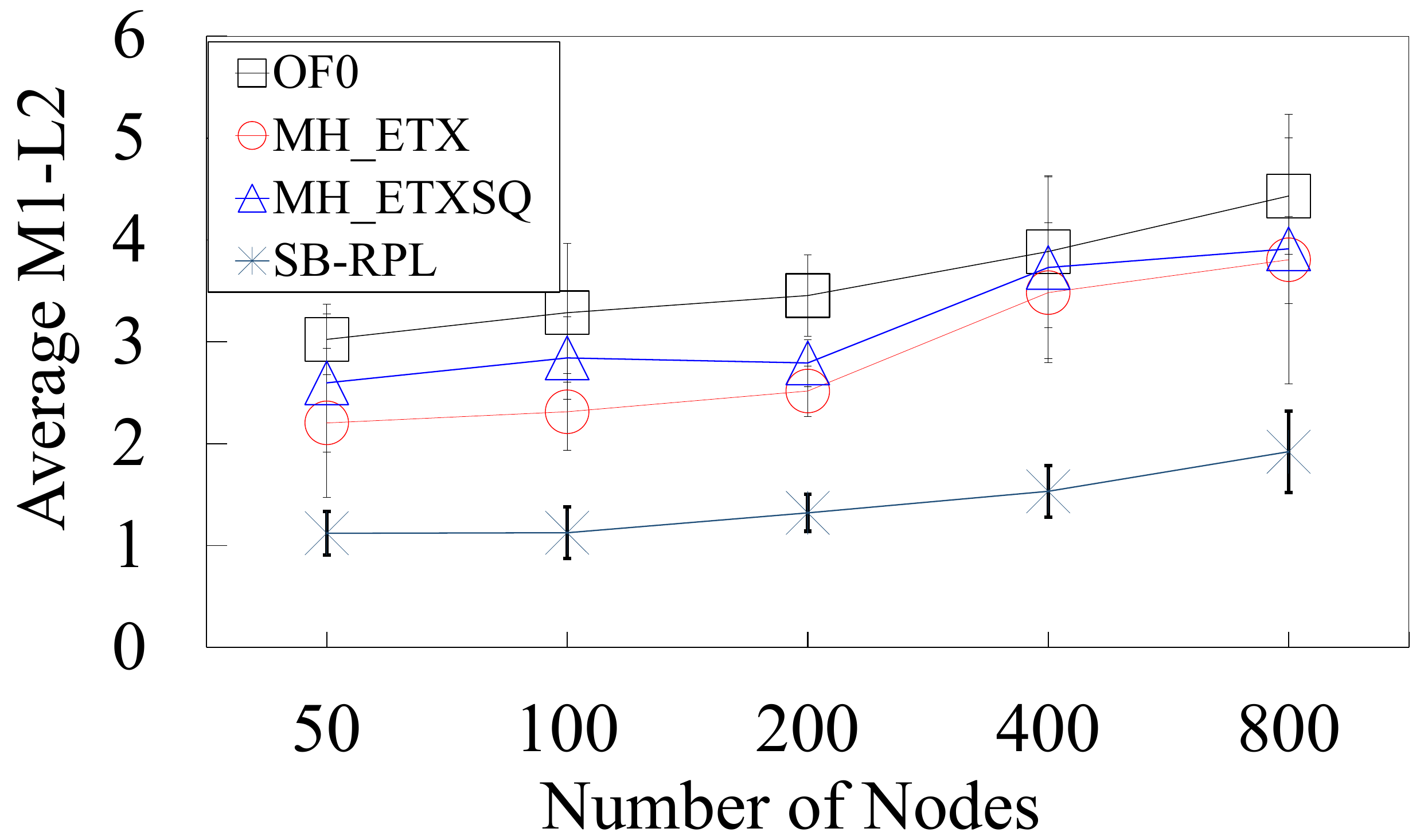}}
	\subcaptionbox{Average $\mathcal{M}2$ - Level 2}{\includegraphics[width=0.24\linewidth, height=0.1\textheight]{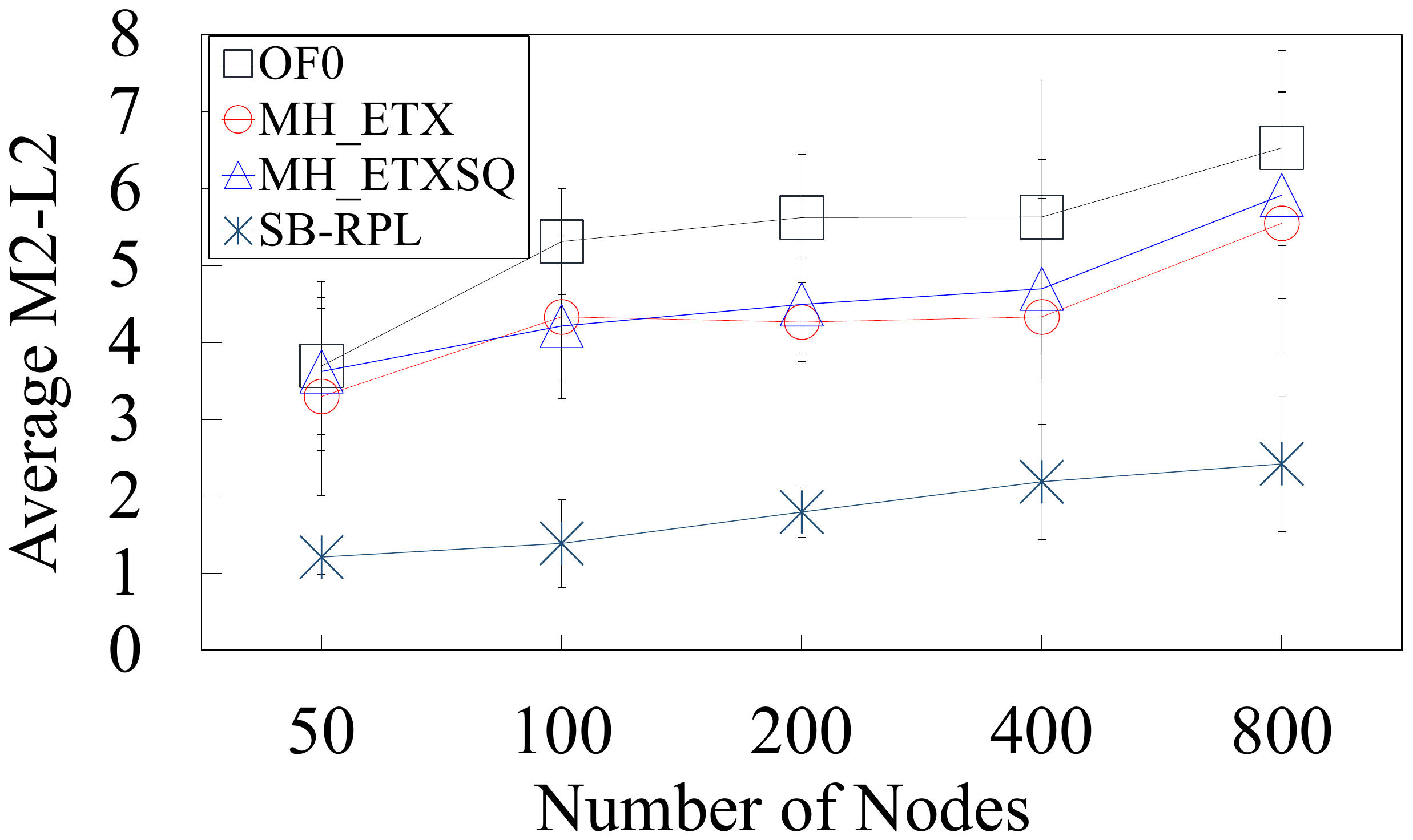}}
	\subcaptionbox{Average $\mathcal{M}3$ - Level 2}{\includegraphics[width=0.24\linewidth, height=0.1\textheight]{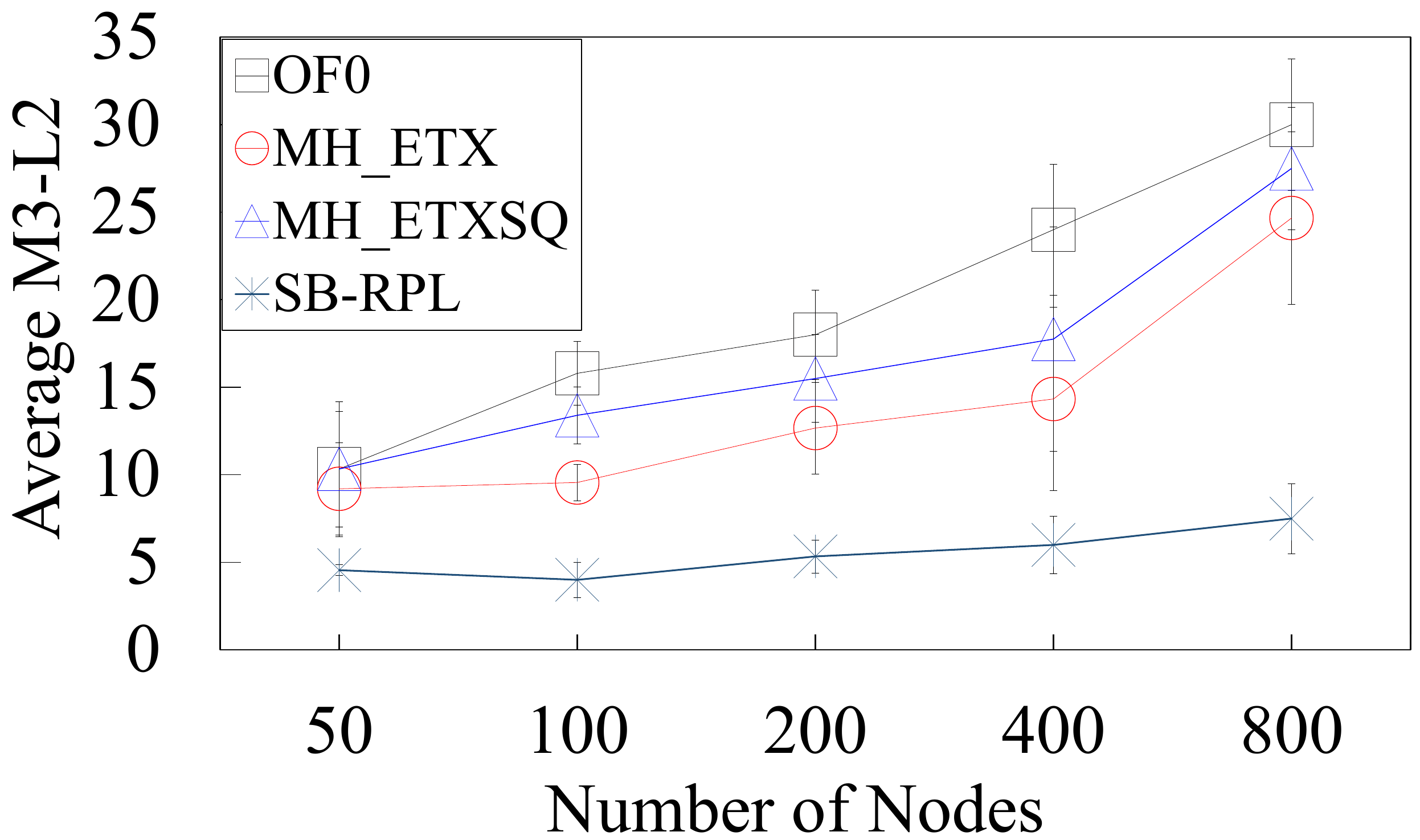}}
	\subcaptionbox{Average $\mathcal{M}4$ - Level 2}{\includegraphics[width=0.24\linewidth, height=0.1\textheight]{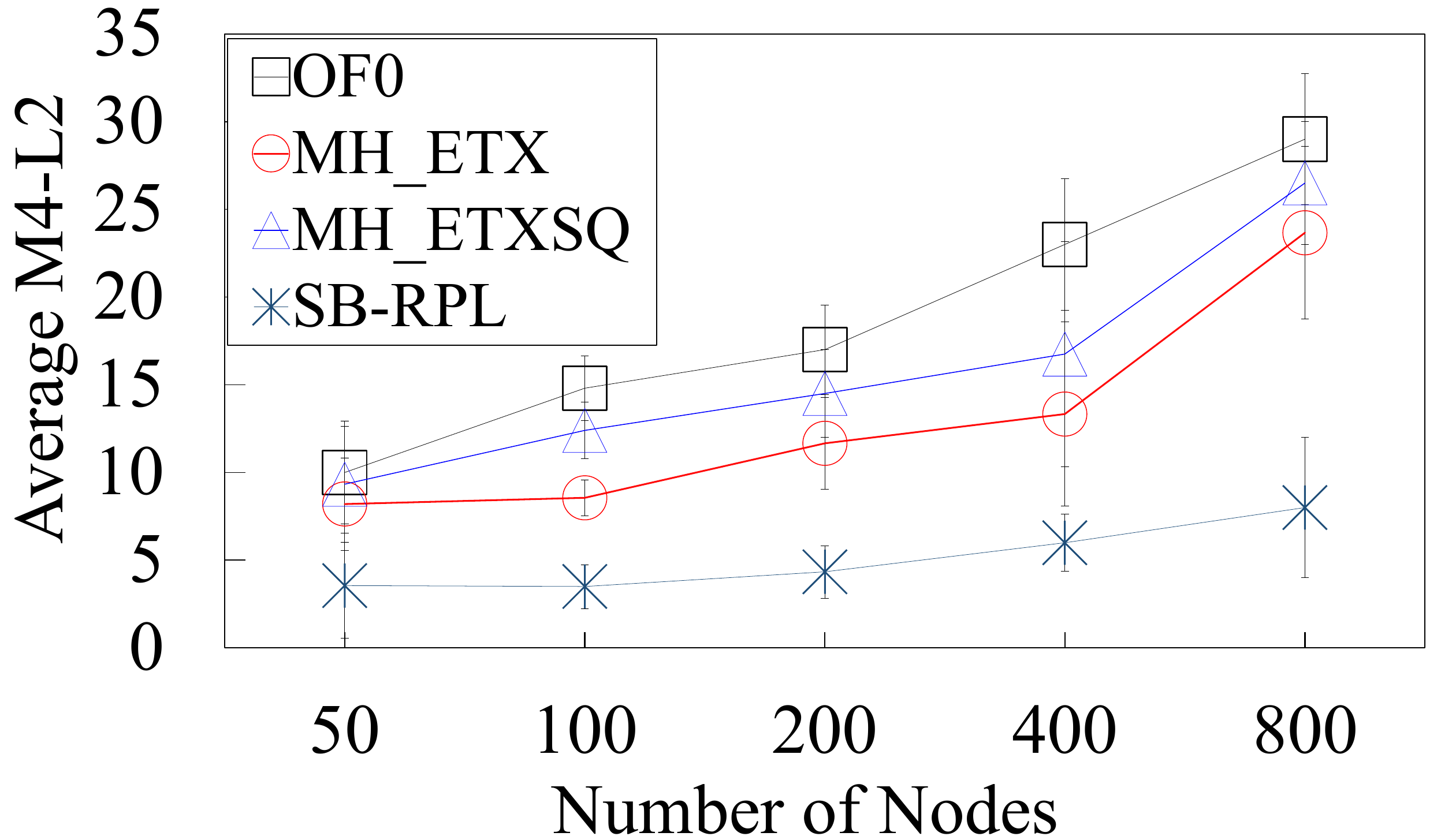}}\\
	\vspace{0.3cm}
	\subcaptionbox{Average $\mathcal{M}1$ - Level 3}{\includegraphics[width=0.24\linewidth, height=0.1\textheight]{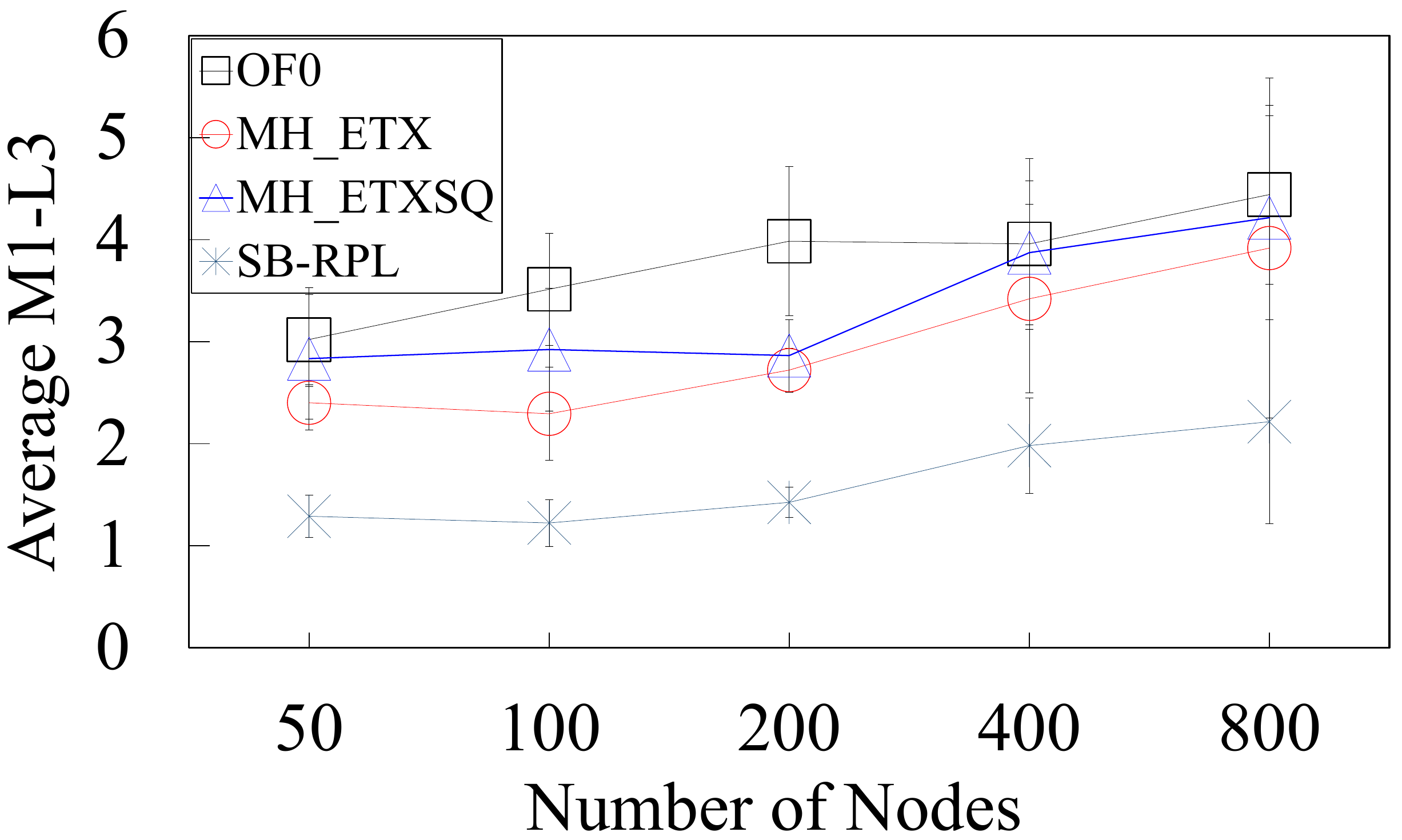}}
	\subcaptionbox{Average $\mathcal{M}2$ - Level 3}{\includegraphics[width=0.24\linewidth, height=0.1\textheight]{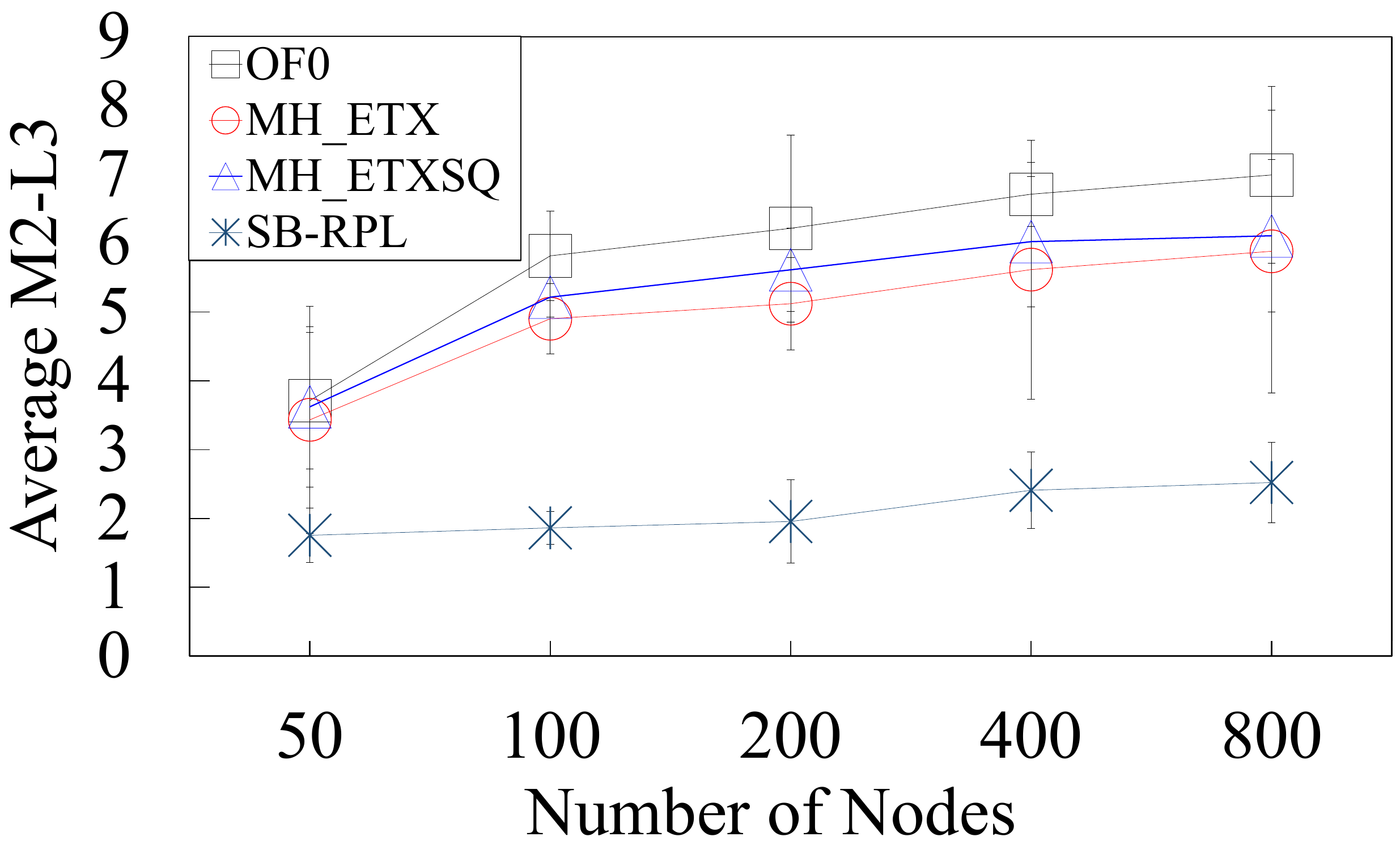}}
	\subcaptionbox{Average $\mathcal{M}3$ - Level 3}{\includegraphics[width=0.24\linewidth, height=0.1\textheight]{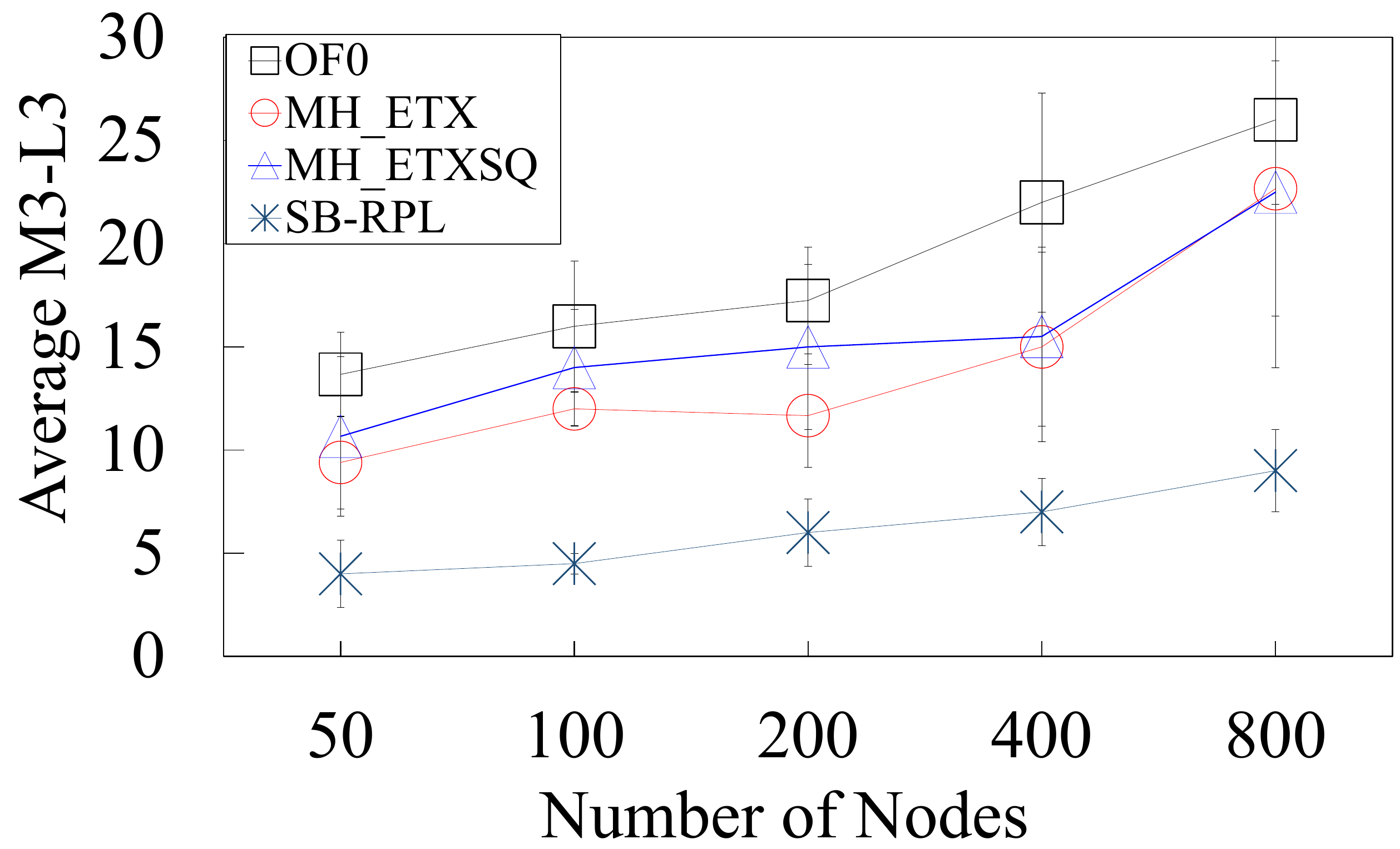}}
	\subcaptionbox{Average $\mathcal{M}4$ - Level 3}{\includegraphics[width=0.24\linewidth, height=0.1\textheight]{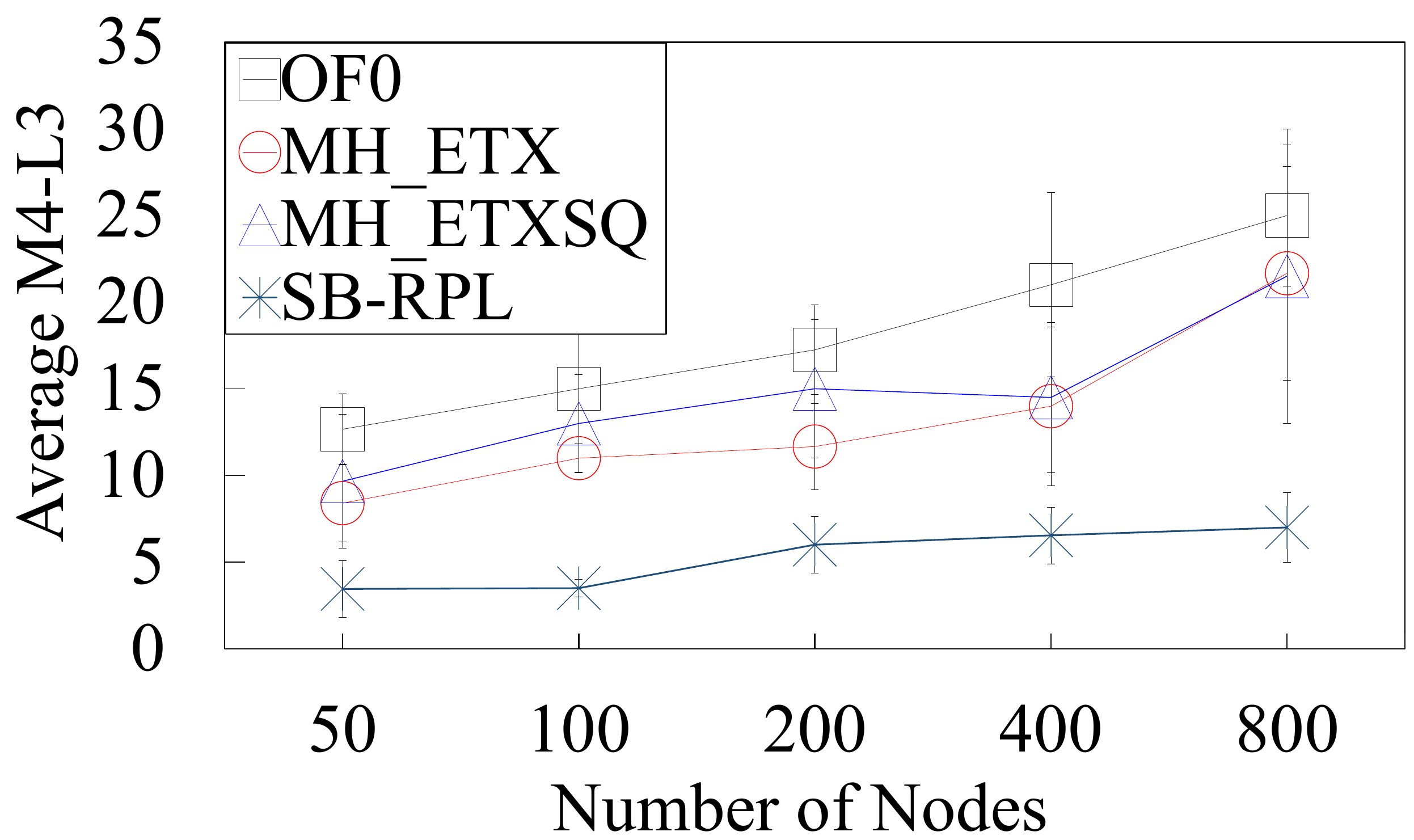}}	
	\caption{ \textbf{Impact of Network Size}. We compare the skewness and balance of evaluated objective functions in different network sizes from 50 nodes to 800 nodes in three levels(1,2,3) of DODAG. The balancing of RPL DODAG topology depends on the network size. Increasing network size leads to reduce topology balance.}
	\label{fig:NWSize}
\end{figure*}%

\subsection{Cooja-based Simulations}
Contiki's network simulator named Cooja is used to provide detail insights about the performance of the proposed scheme under other network conditions. First, these simulations compare the performance of the proposed scheme for a network following various network sizes with 50, 100, 200, 400 and 800 nodes. Then, we demonstrate that the proposed scheme is outstanding performance under low, medium and  dense networks compared to prior approaches.%
 
\subsubsection{Impact of Network Scales}
Fig.~\ref{fig:NWSize} illustrates  the impact of network sizes (50,100, 200, 400, and 800 nodes) to skewness indexes of RPL DODAG topology. As is observed from the graph, the skewness indexes including $\mathcal{M}1$, $\mathcal{M}2$, $\mathcal{M}3$, and $\mathcal{M}4$ in three separate levels (1,2,3) of existing routing strategies are higher almost three times our proposed scheme. Overall, the skewness values $\mathcal{M}1$, $\mathcal{M}2$, $\mathcal{M}3$, and $\mathcal{M}4$ of SB-RPL are significantly lower than the existing objective functions such as OF0, MRHOF. As the increase of the number of nodes, the skewness indexes of OF0 and MRHOF also raise remarkably while the ones of SB-RPL go up slightly. From Fig.~\ref{fig:NWSize}(a), Fig.~\ref{fig:NWSize}(b), Fig.~\ref{fig:NWSize}(c) and Fig.~\ref{fig:NWSize}(d), we compare the average values of $\mathcal{M}1$, $\mathcal{M}2$, $\mathcal{M}3$, and $\mathcal{M}4$ at the same Level 1. The skewness indexes $\mathcal{M}1$, $\mathcal{M}2$, $\mathcal{M}3$, and $\mathcal{M}4$ of RPL-OF0, RPL-MH$\_$ETX, RPL-MH$\_$ETXSQ are quite similar and much higher than the ones of SB-RPL. Even when the network size is large as 400-node networks and 800-node networks, the balancing within SB-RPL DODAG is still quite stable. The reason for this is because SB-RPL considers the skewness and balanced metric in parent selection procedure to balance the size among subtrees in RPL.%

\begin{figure*}[t!]
	\centering
	\frame{\includegraphics[width=0.3\linewidth, height=0.02\textheight]{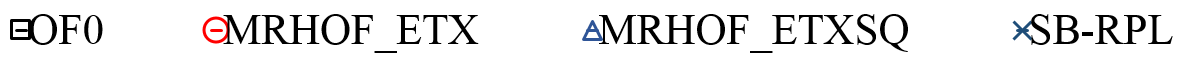}} \\
	\subcaptionbox{Average $\mathcal{M}1$ - Level 1}{\includegraphics[width=0.24\linewidth, height=0.1\textheight]{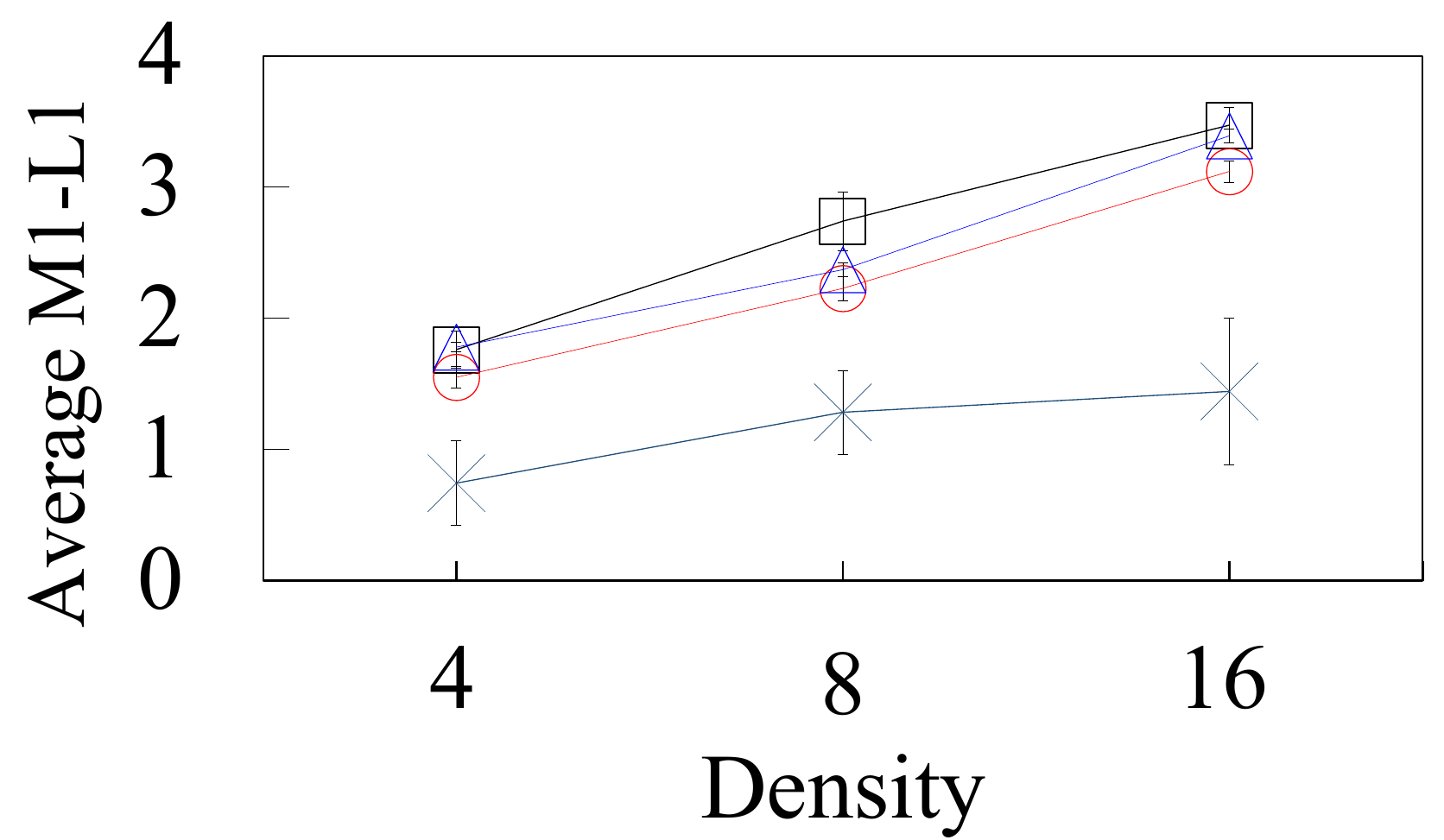}}
	\subcaptionbox{Average $\mathcal{M}2$ - Level 1}{\includegraphics[width=0.24\linewidth, height=0.1\textheight]{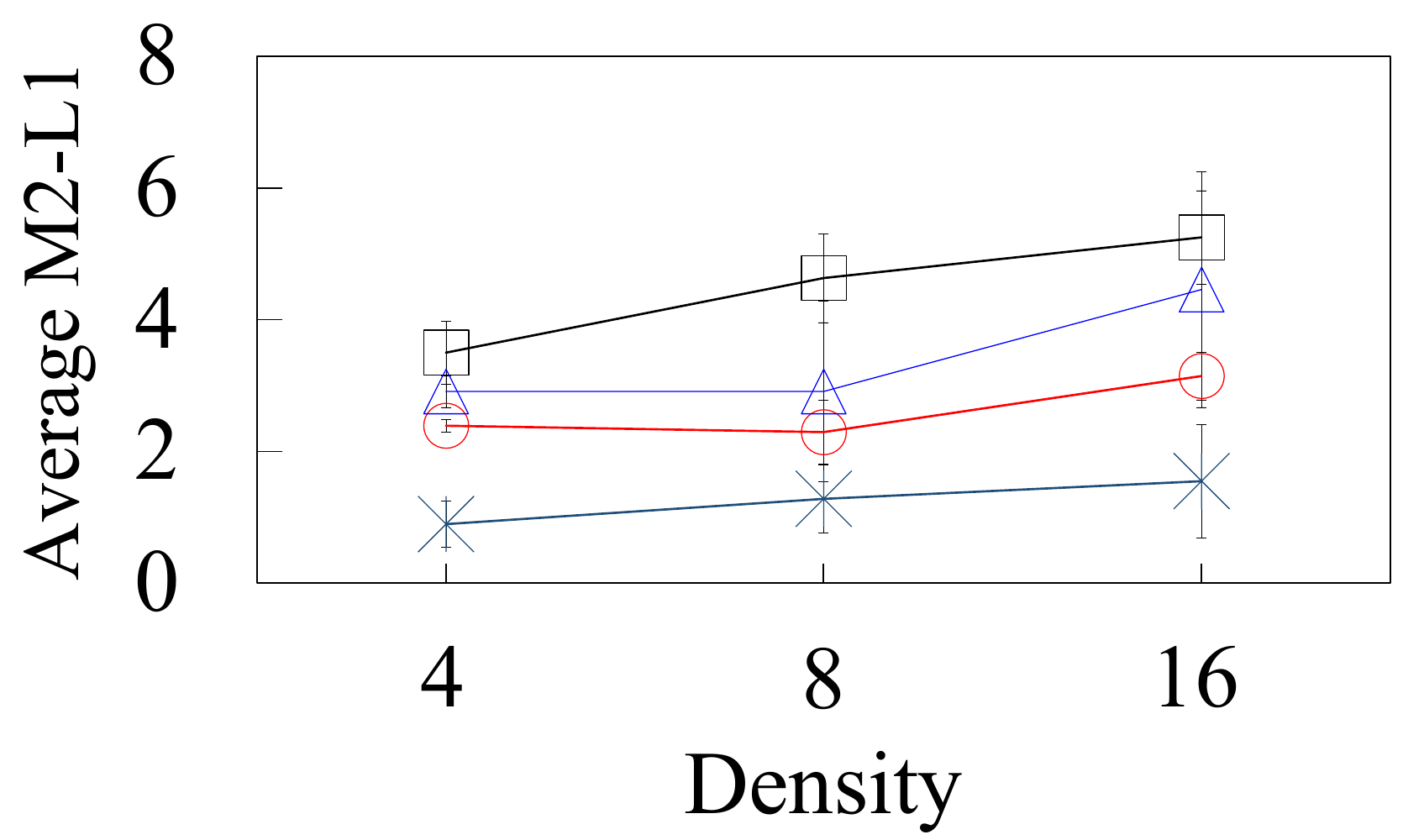}}
	\subcaptionbox{Average $\mathcal{M}3$ - Level 1}{\includegraphics[width=0.24\linewidth, height=0.1\textheight]{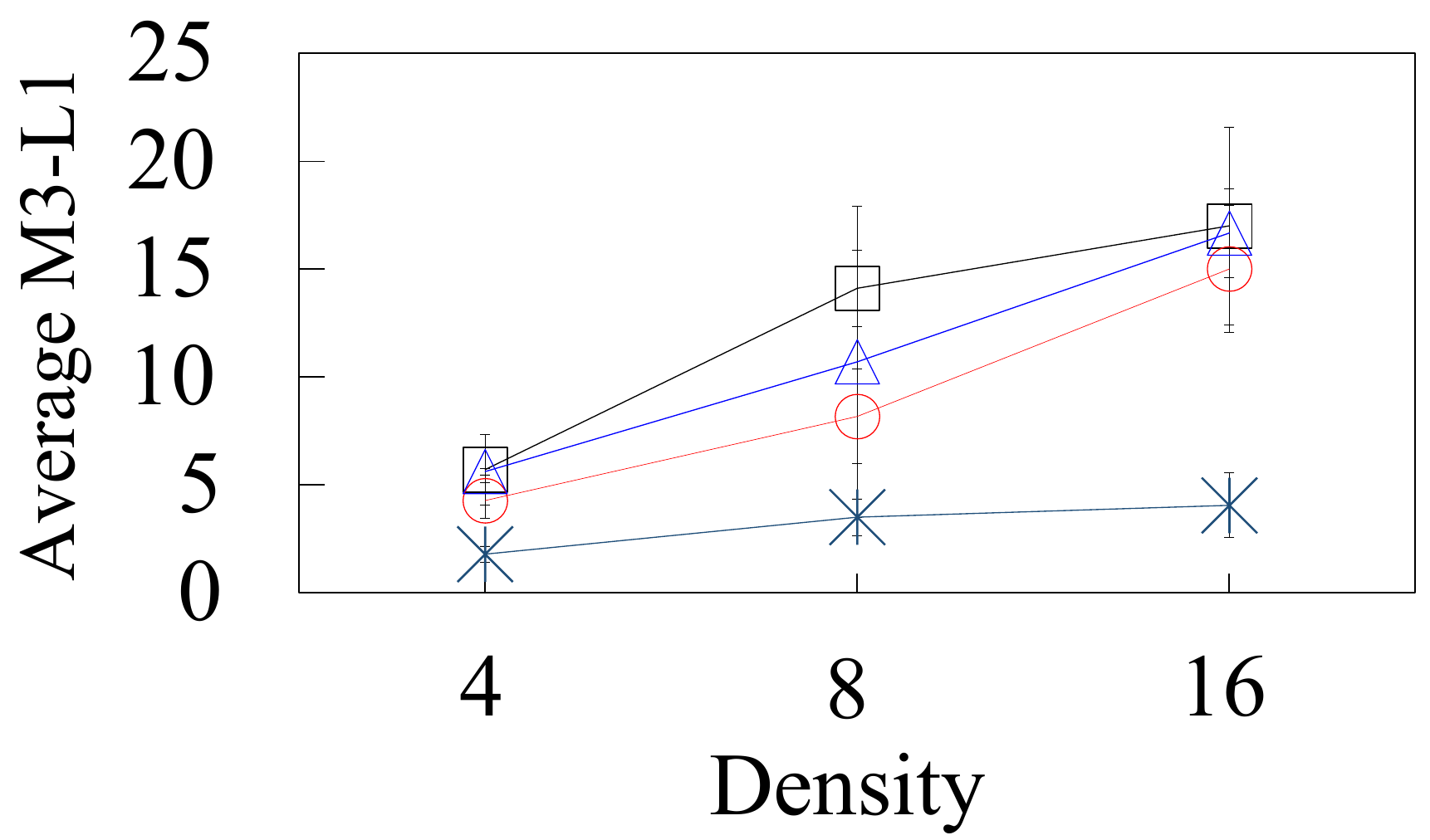}}
	\subcaptionbox{Average $\mathcal{M}4$ - Level 1}{\includegraphics[width=0.24\linewidth, height=0.1\textheight]{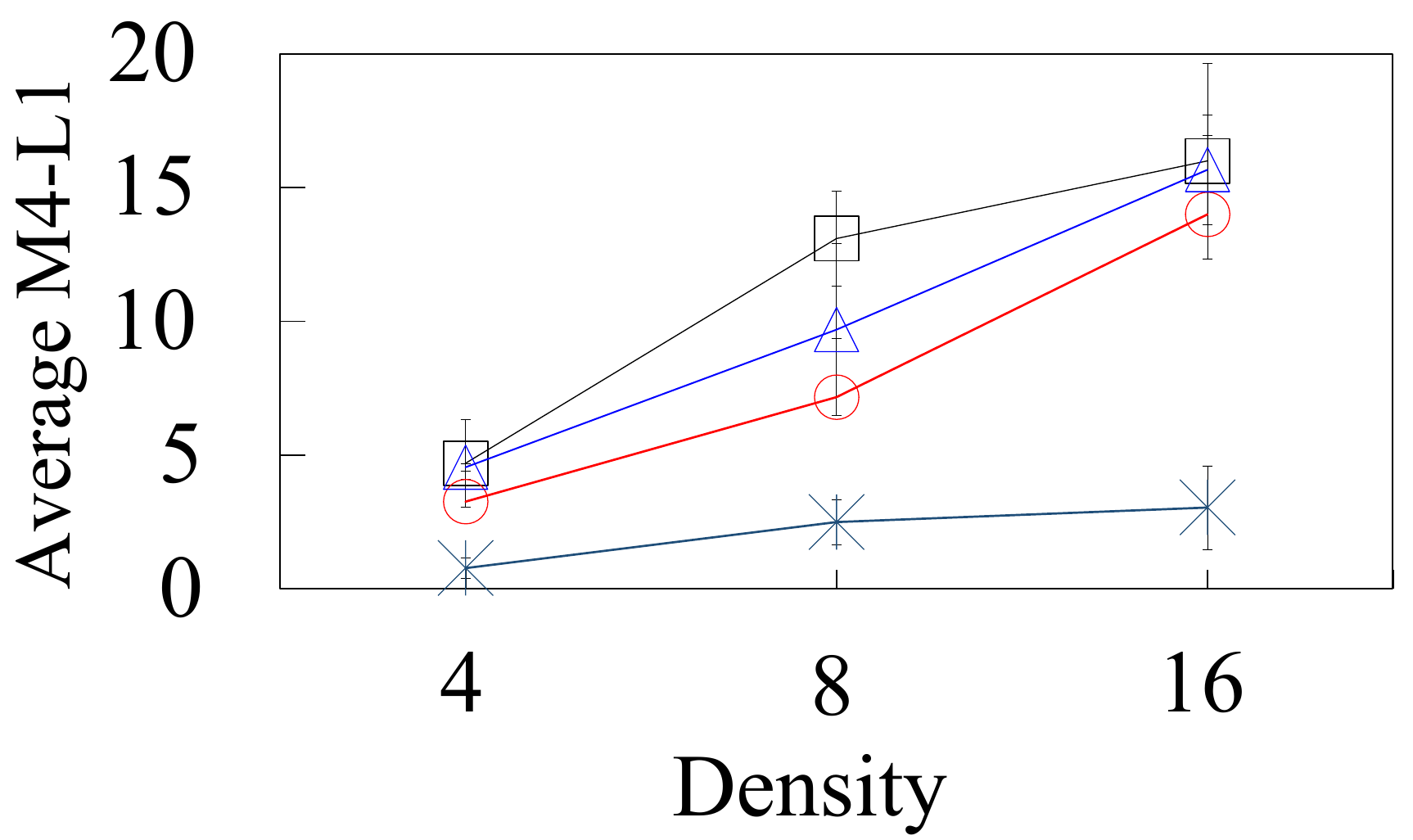}}\\
	\vspace{0.3cm}
	\subcaptionbox{Average $\mathcal{M}1$ - Level 2}{\includegraphics[width=0.24\linewidth, height=0.1\textheight]{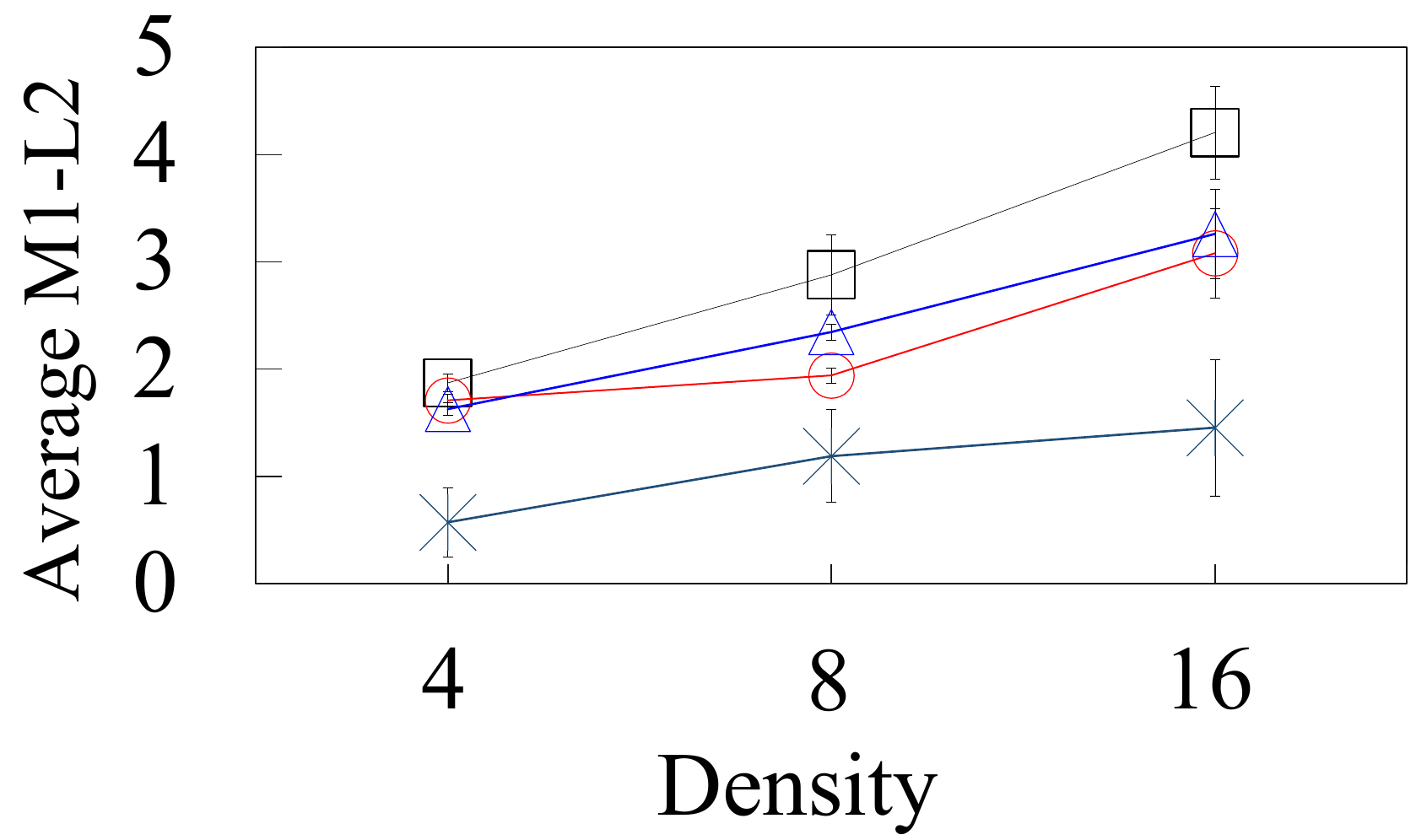}}
	\subcaptionbox{Average $\mathcal{M}2$ - Level 2}{\includegraphics[width=0.24\linewidth, height=0.1\textheight]{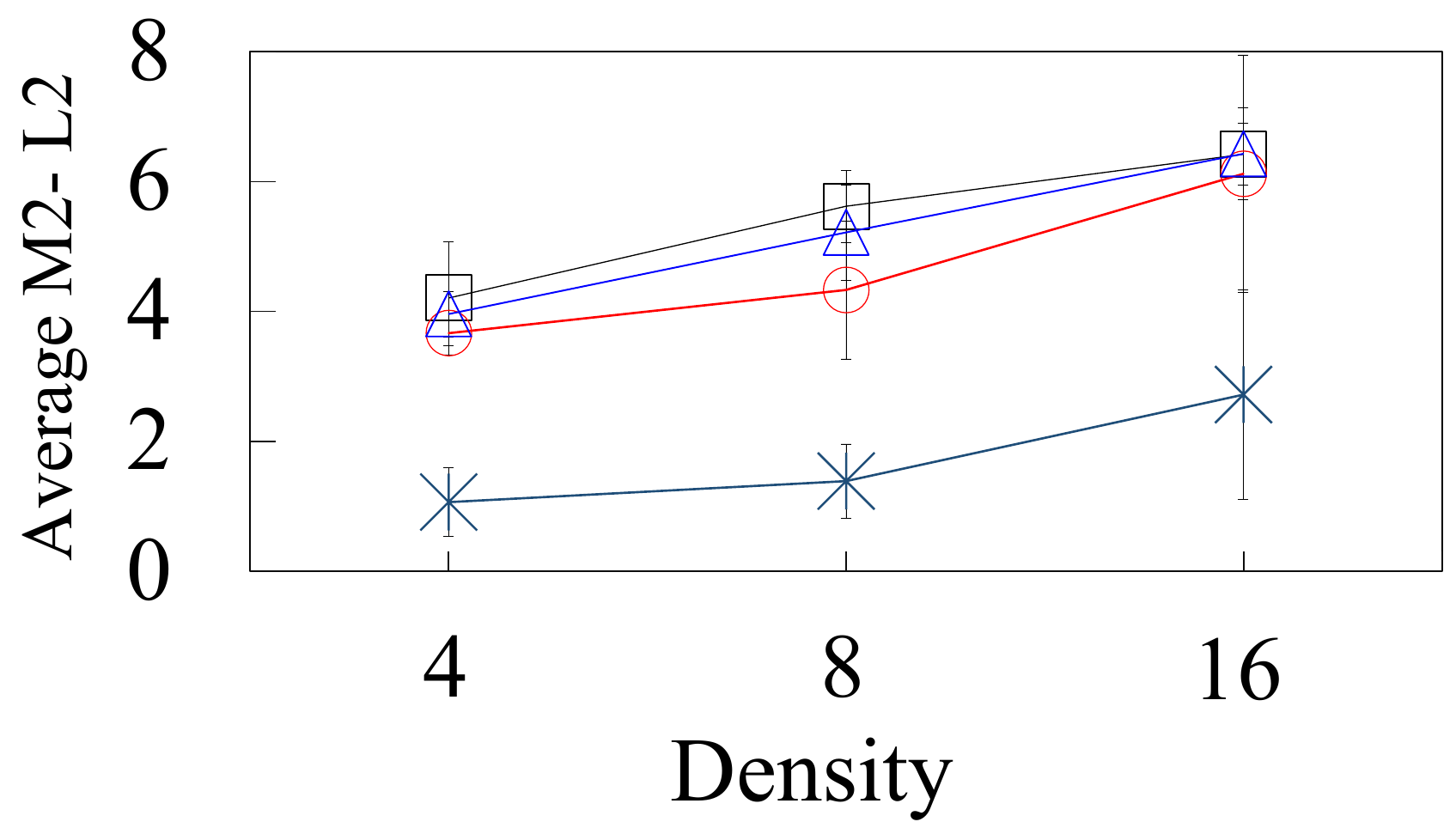}}
	\subcaptionbox{Average $\mathcal{M}3$ - Level 2}{\includegraphics[width=0.24\linewidth, height=0.1\textheight]{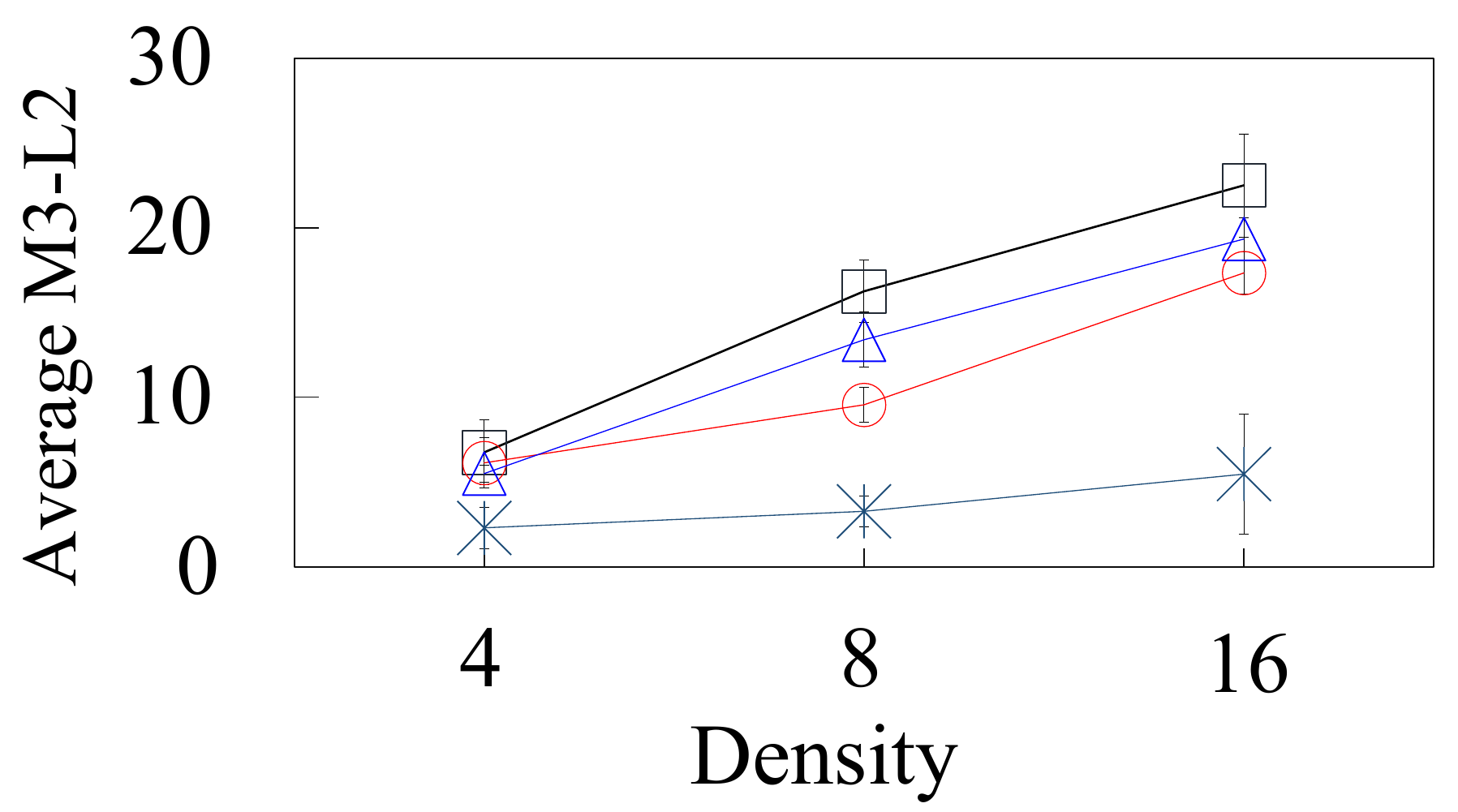}}
	\subcaptionbox{Average $\mathcal{M}4$ - Level 2}{\includegraphics[width=0.24\linewidth, height=0.1\textheight]{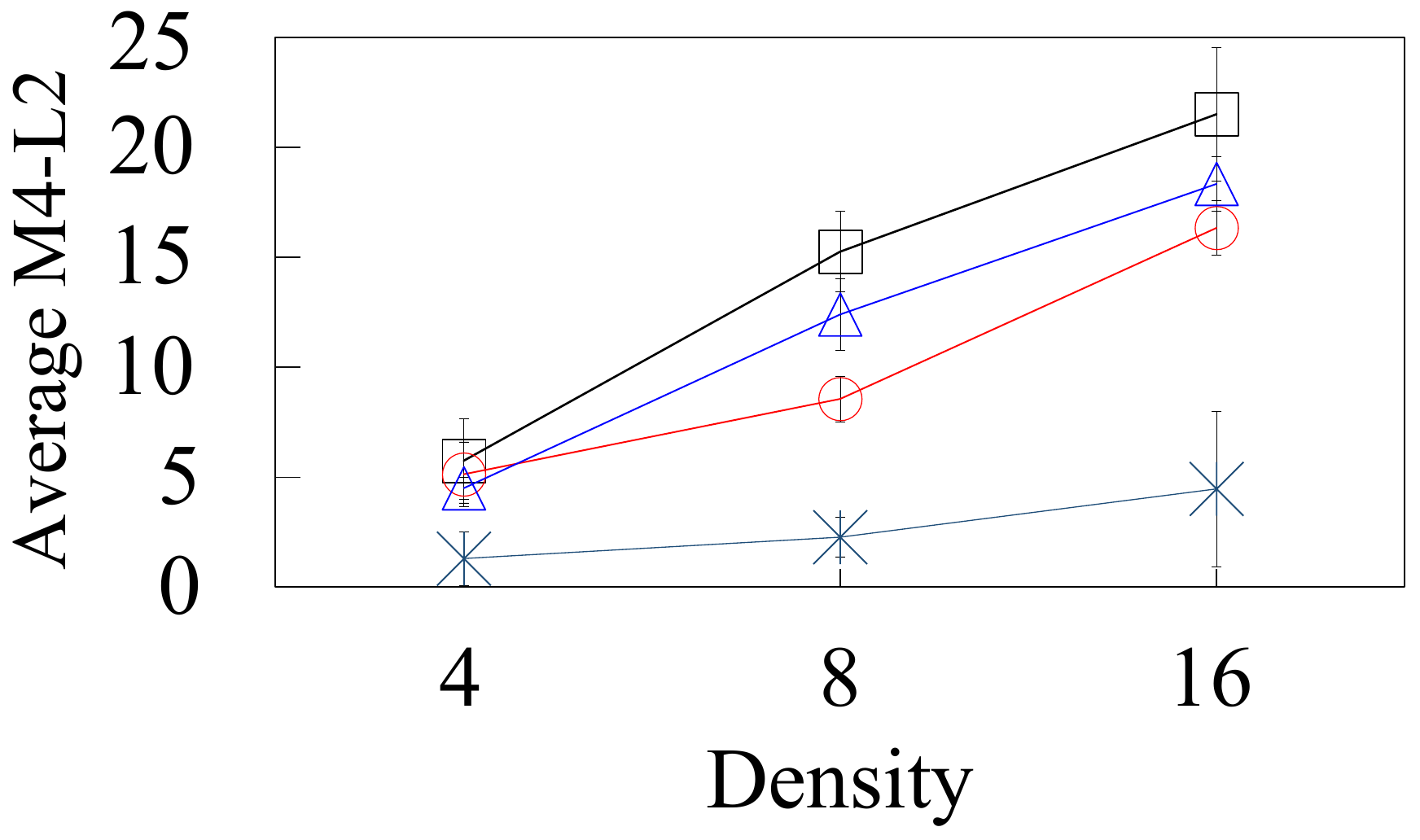}} \\
	\vspace{0.3cm}
	\subcaptionbox{Average $\mathcal{M}1$ - Level 3}{\includegraphics[width=0.24\linewidth, height=0.1\textheight]{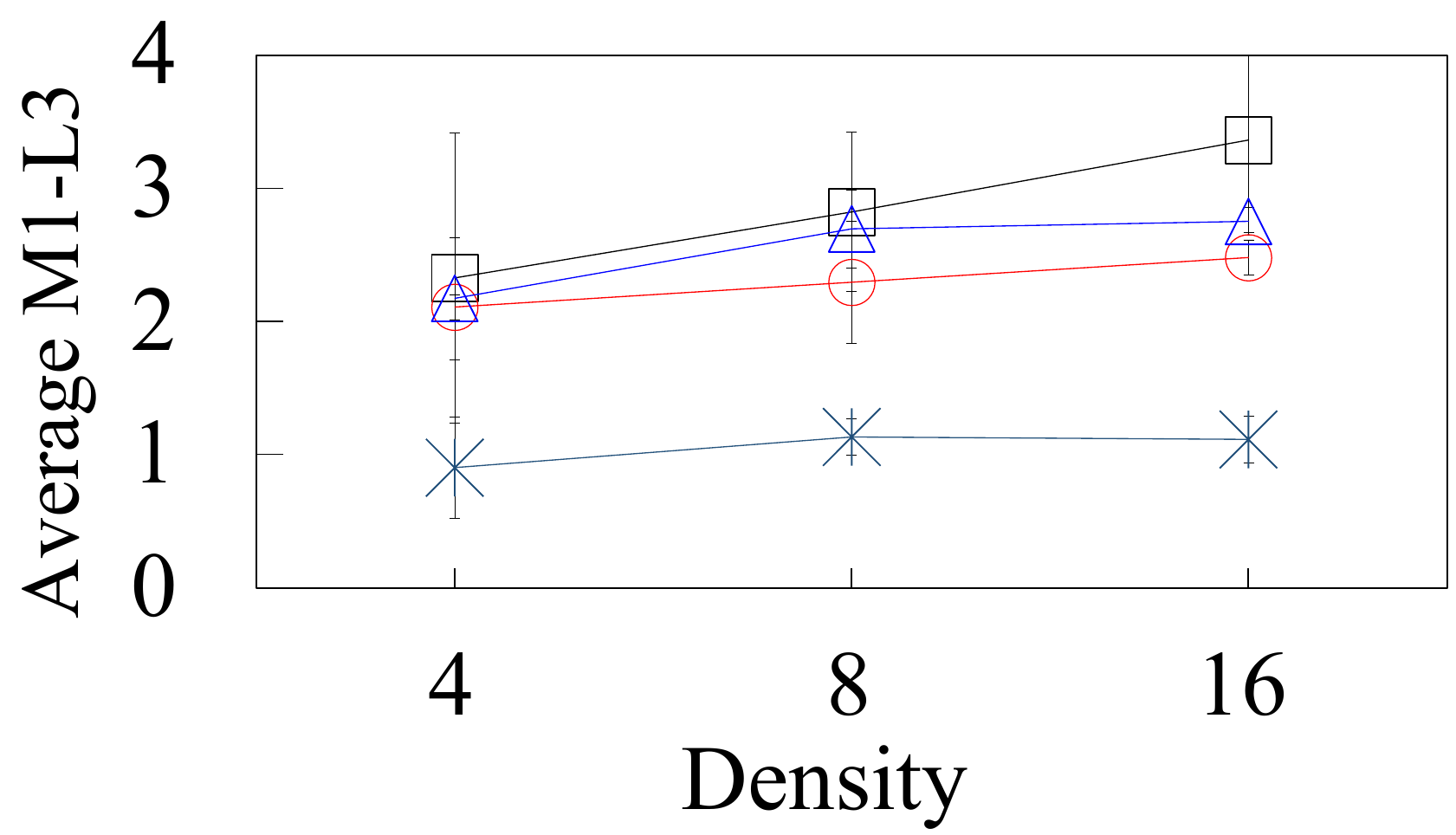}}
	\subcaptionbox{Average $\mathcal{M}2$ - Level 3}{\includegraphics[width=0.24\linewidth, height=0.1\textheight]{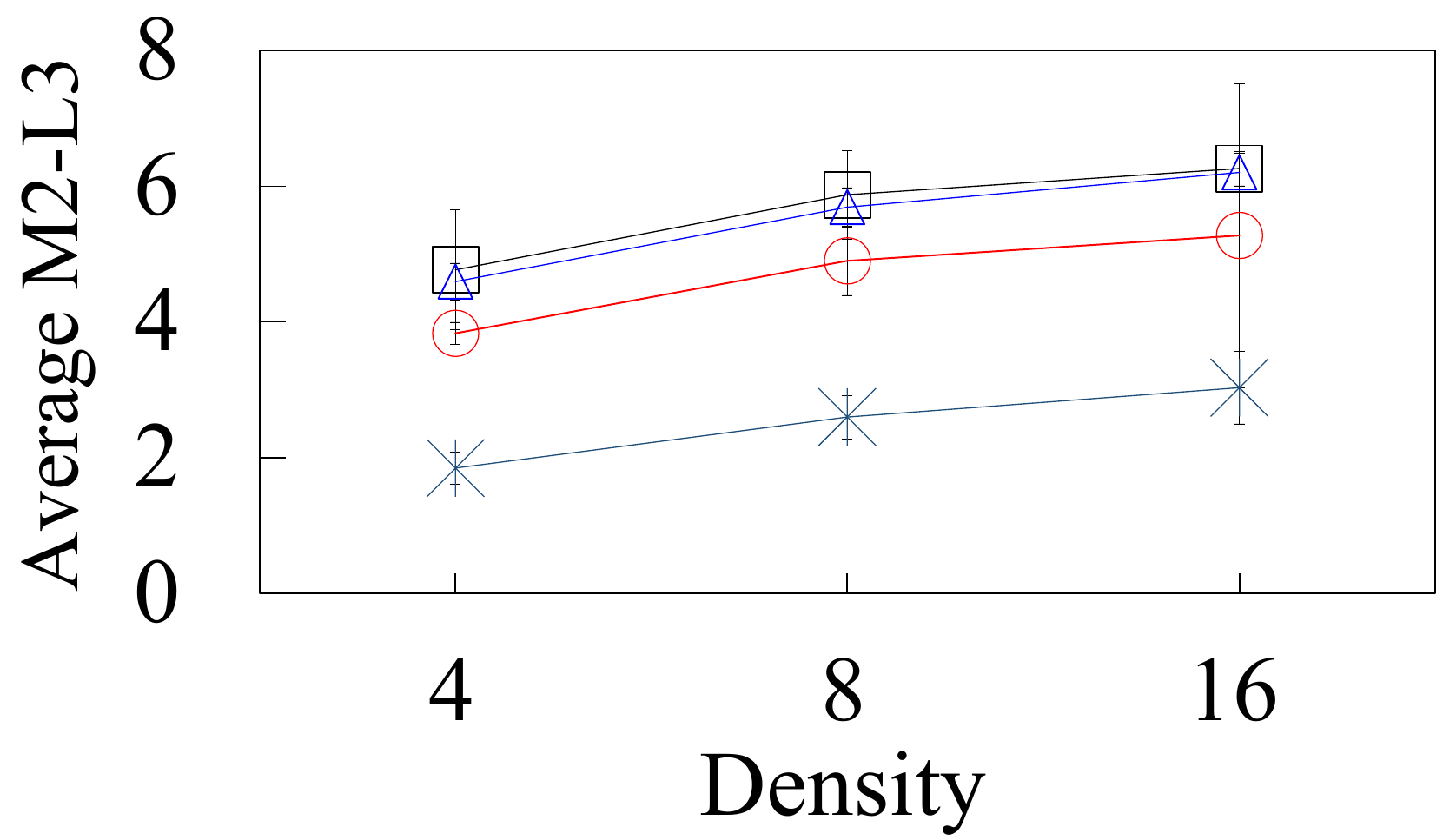}}
	\subcaptionbox{Average $\mathcal{M}3$ - Level 3}{\includegraphics[width=0.24\linewidth, height=0.1\textheight]{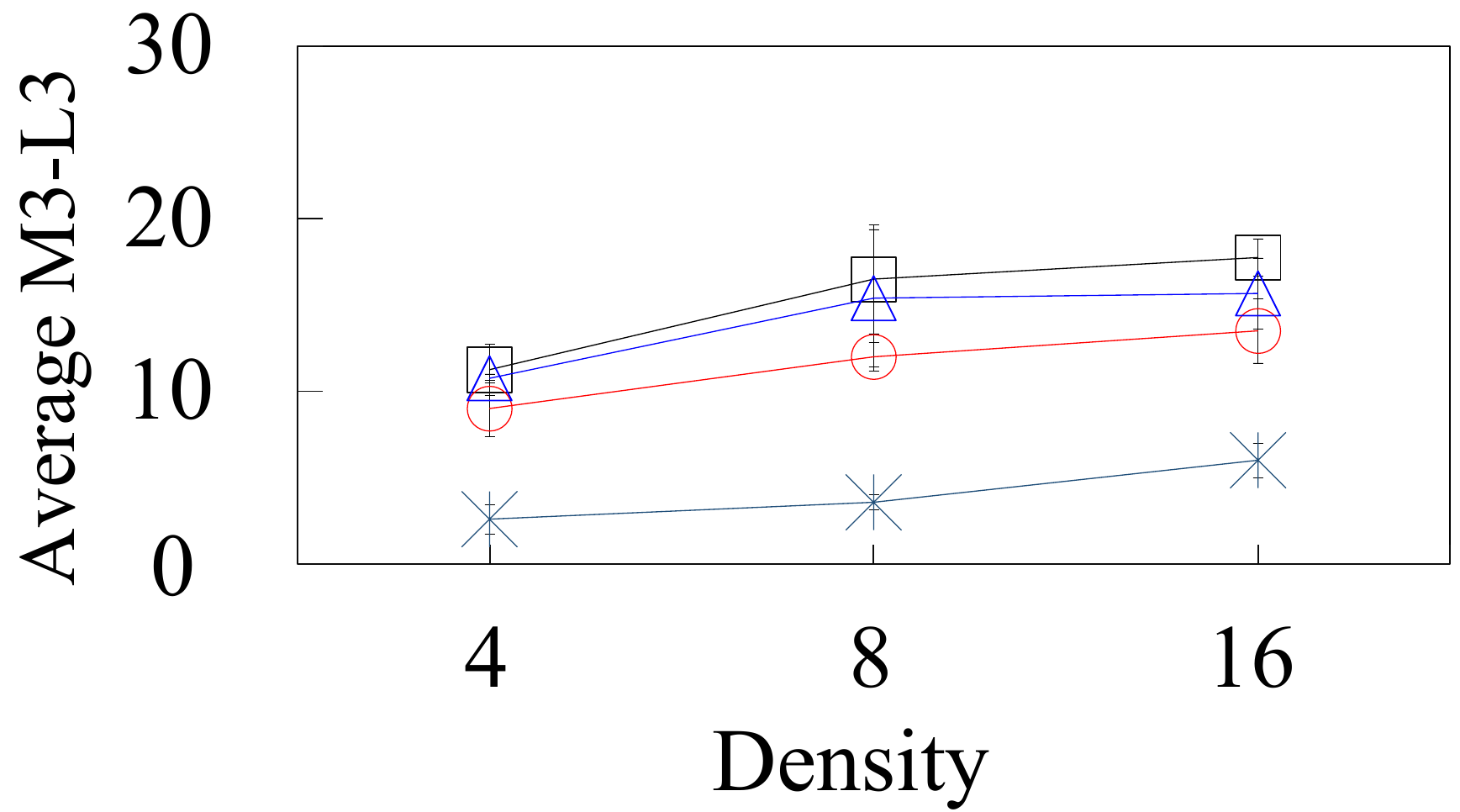}}
	\subcaptionbox{Average $\mathcal{M}4$ - Level 3}{\includegraphics[width=0.24\linewidth, height=0.1\textheight]{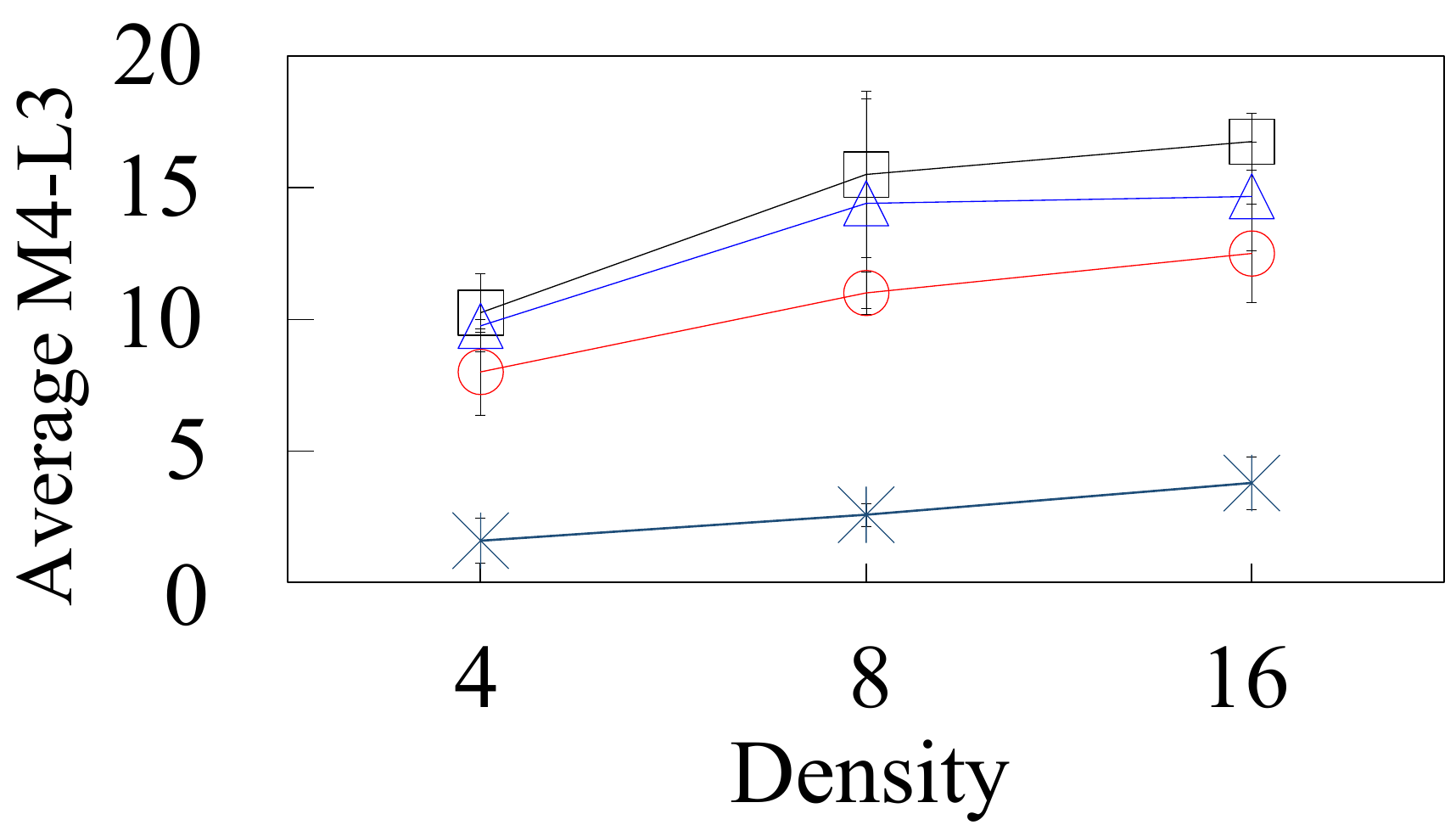}}
	\caption{\textbf{Impact of Network Density}. The balancing of RPL DODAG topology depends on the network size. With traditional objective functions such as OF0 and MRHOF, the skewness of RPL increases exponentially as the growth of network density while the proposed scheme keeps the RPL topology stably.}
	\label{fig:NwDensity}
\end{figure*}%

\subsubsection{Impact of Network Densities}
One of the challenges which effects the stable of DODAG is network density. As the number of sensor node increases, each node will have more connections with neighbors. Fig.~\ref{fig:NwDensity} compares the average skewness indexes under various types of topologies with different densities 4, 8 and 16 which equivalent to low, medium, and dense networks respectively. Overall, the skewness indexes increase exponentially as the increase of density. Fig.~\ref{fig:NwDensity}(a), Fig.~\ref{fig:NwDensity}(b), Fig.~\ref{fig:NwDensity}(c), and Fig.~\ref{fig:NwDensity}(d) compare the averages skewness indexes in level 1 of DODAG. The average value of skewness indexes increases rapidly as the increment of density in low and medium density. Then the skewness indexes increase slightly when the density increases from 8 to 16. At level 1, in dense networks, the skewness indexes show the big gap between SB-RPL and other schemes. At level 2 and level 2, the nodes are distributed widely, the skewness indexes of OF0, MH$\_$ETX, and MH$\_$ETXSQ still rise gradually. Meanwhile, the $\mathcal{M}1$, $\mathcal{M}2$, $\mathcal{M}3$, and $\mathcal{M}4$ of SB-RPL keep stably in various types of networks. When the network becomes denser, all existing routing schemes perform significantly better in terms of end-to-end latency, meanwhile slightly worse in terms of packet loss ratio and overhead. However, SB-RPL significantly outperforms other protocols. First, thanks to $ST_p(t)$ metric which indicates the size of the subtree, the skewness among subtrees in DODAG  remains stably and avoids traffic congestion for thoroughly topologies. Second, besides \textit{Subtree Size}, the efficiency of routing procedure relies on considering the link quality from joining node and parent candidate to guarantee quality for data transmission, remaining the stable for RPL. Consequently, the skewness indexes of SB-RPL is always less than 3 times to existing schemes in low, medium, and dense network environments. 

\section{Conclusion}
In conclusion, we discussed the load balancing problem which is the key issue but it has not addressed efficiently when designing objective functions for RPL, given that scalability, energy efficiency and resource constrained are main characteristics of Low-Power and Lossy Networks. In effect, the load imbalance problems of RPL decrease the performance as well as waste network resources. 

To remedy these problems, we proposed a light-weight but effective solution, called SB-RPL, that aims to achieve balanced workload distribution among nodes in large-scale low power and lossy networks by exploiting the combination of multiple routing metrics as well as the skewness and balance among subtrees in RPL DODAG in support routing procedure. 

We implemented SB-RPL in ContikiOS and conducted an extensive evaluation using computer simulation and on large-scale real-world testbed. We demonstrated that the practicality of SB-RPL and its ability to consistently achieve the great balancing RPL trees and high end-to-end packet delivery performance by alleviating the congestion and providing the ability to support large networks. 

As a part of future work, we are studying resource fairness issues among multiple RPL DODAGs and we intend to define a SDN-based\cite{icoin} mechanism in support RPL routing operation.%   

\end{document}